\documentclass[aps,twocolumn,showpacs,amsmath,amssymb,prl,superscriptaddress,floatfix]{revtex4-1}
\usepackage{graphicx}
\usepackage{bm}
\usepackage{bbold}

\usepackage{verbatim}
\usepackage{amsmath,amsfonts,amssymb}
\usepackage{color}

\usepackage{dsfont}

\usepackage{multibib}
\usepackage{hyperref}

\renewcommand{\L}{{\mathcal L}}
\newcommand{\M}{{\mathcal M}}
\newcommand{\Leff}{{\mathcal L}_{\rm eff}}
\renewcommand{\H}{{\mathcal H}}
\newcommand{\K}{{\mathcal K}}
\renewcommand{\P}{{\mathcal P}}
\newcommand{\Q}{{\mathcal Q}}
\newcommand{\I}{{\mathcal I}}

\renewcommand{\S}{{\mathcal S}}
\newcommand{\rt}{{\rho(t)}}
\newcommand{\rss}{{\rho_{\rm ss}}}
\newcommand{\ri}{{\rho_{\rm in}}}

\newcommand{\rms}{{\rho_{\rm MS}}}
\renewcommand{\r}{{{R }}}
\renewcommand{\l}{{{L }}}
\newcommand{\Tr}{{\mathrm{Tr}}}
\newcommand{\tr}{{\widetilde{\rho}}}
\newcommand{\tp}{{\widetilde{P}}}
\renewcommand{\dim}{\mathrm{dim}}

\definecolor{green}{rgb}{0.0, 0.5, 0.0}

\renewcommand{\Re}{{{\mathrm{Re}}}}
\renewcommand{\Im}{{{\mathrm{Im}}}}

\begin{document}

\title{Towards a theory of metastability in open quantum dynamics}

\author{Katarzyna Macieszczak}
\affiliation{School of Mathematical Sciences, University of Nottingham, Nottingham, NG7 2RD, UK}
\affiliation{School of Physics and Astronomy, University of Nottingham, Nottingham, NG7 2RD, UK}
\author{M\u{a}d\u{a}lin Gu\c{t}\u{a}}
\affiliation{School of Mathematical Sciences, University of Nottingham, Nottingham, NG7 2RD, UK}
\author{Igor Lesanovsky}
\author{Juan P. Garrahan}
\affiliation{School of Physics and Astronomy, University of Nottingham, Nottingham, NG7 2RD, UK}


\date{\today}

\begin{abstract}
By generalising concepts from classical stochastic dynamics, we establish the basis for a theory of metastability in Markovian open quantum systems.  Partial relaxation into long-lived metastable states---distinct from the asymptotic stationary state---is a manifestation of a separation of timescales due to a splitting in the spectrum of the generator of the dynamics.  We show here how to exploit this spectral structure to obtain a low dimensional approximation to the dynamics in terms of motion in a manifold of metastable states constructed from the low-lying eigenmatrices of the generator.  We argue that the metastable manifold is in general composed of disjoint states, noiseless subsystems and decoherence-free subspaces.  \end{abstract}

\maketitle

\noindent
{\bf \em Introduction.} Stochastic many-body systems often display complex and slow relaxation towards a stationary state.  A common phenomenon is that of {\em metastability}, where initial relaxation is into long-lived states, with subsequent decay to true stationarity occurring at much longer times. This separation of times in the dynamics has evident experimental manifestations, for example in two-step decay of time correlation functions.  Metastability is a common occurrence in classical soft matter \cite{Binder2011}, glasses being the paradigmatic example \cite{Biroli2013,Berthier2015}.

There is much current interest in the non-equilibrium dynamics of quantum many-body systems, both closed (i.e., isolated) and open (i.e., interacting with an environment).  This includes issues such as thermalisation \cite{Polkovnikov2011,Eisert2014,Dalessio2015,Gogolin2015}, many-body localisation \cite{Nandkishore2015,Yao2014,De-Roeck2014}, and aging and glassy behaviour, where questions about timescales and partial versus full relaxation play central roles \cite{Prosen2011,Markland2011,Olmos2012,Sciolla2015,Horssen2015,Znidaric2015}.  
From the quantum information perspective, decoherence free subspaces  \cite{Zanardi1997d,Zanardi1997e,Lidar1998c,Kielpinski2001b} and noiseless subsystems \cite{Knill2000b,Zanardi2001b,Viola2001b}, where parts of the Hilbert space are protected against external noise, are ideal scenarios for implementing quantum information processing \cite{Nielsen2000}.  Since experiments are performed in finite time, it is sufficient (and practical) to consider manifolds of coherent states which are only stable over experimental timescales, i.e., metastable, with respect to noise.

Given this broad range of problems, it would be highly desirable to have a unified theory of quantum metastability.  In this paper we lay the ground for such a theory for the case of open quantum systems evolving with Markovian dynamics.  Our starting point is a well-established approach for metastability in classical stochastic systems \cite{Gaveau1996,*Gaveau1998,*Gaveau1999,Bovier2002,Gaveau2006,Nicholson2013,[For a pedagogical review see: ]Kurchan2009}.  
We develop an analogous method for quantum Markovian systems based on the spectral properties of the generator of the dynamics.  Separation of timescales implies a splitting in the spectrum, and this spectral division allows us to construct metastable states from the low-lying eigenmatrices of the generator.  
Based on perturbative calculations for finite systems, we argue 
that the manifold of metastable states is in general composed of disjoint states, noiseless subsystems and decoherence-free subspaces.  
We illustrate these possibilities with simple examples.  We further discuss how to reduce the overall dynamics to a low-dimensional effective motion in the metastable manifold, and consider the associated behaviour of time correlations.

\smallskip

\noindent
{\bf \em Quantum metastability and spectral properties.} We consider an open quantum system evolving under Markovian dynamics, with Linbladian master equation $\frac{d}{dt}\rho(t) = \mathcal{L} \rho(t)$ \cite{Lindblad1976,Gorini1976,Plenio1998,Gardiner2004}, where the generator of the dynamics $\L$ is,
\begin{equation}
\mathcal{L} (\cdot) := -i[H,(\cdot)] + \sum_{j}
\left(
J_{j} (\cdot) J_{j}^\dagger - \frac{1}{2}\{J_{j}^\dagger J_{j} , (\cdot)\} 
\right) .
\label{eq:master}
\end{equation}
The state of the system at time $t$ is $\rho(t)$, the system Hamiltonian is $H$, and $\{ J_j \}$ are
quantum jump operators \footnote{
Calligraphic font denotes super-operators, such as the generator $\L$, 
while Roman font denotes normal operators, such as the Hamiltonian $H$ or the jump operators $J_{i}$. 
}.  While in general the linear operator $\L$ is not diagonalisable, one can find its eigenvalues $\{ \lambda_{k}, k=1,2,\ldots\}$ [which we order by decreasing real part, $\Re(\lambda_{k}) \geq \Re(\lambda_{k+1})$] each corresponding to an eigenspace or a Jordan block.  Since $\L$ generates a proper quantum stochastic (completely positive trace-preserving) dynamics of $\rt$, its largest eigenvalue vanishes, $\lambda_1=0$, and its associated right eigenmatrix $\r_{1}$ is the stationary state, $\r_{1}=\rss$ (the corresponding left eigenmatrix being the identity, $\l_{1}=I$) 
\footnote{
$\r_{k}$ and $\l_{k}$ are right and left eigenmatrices of $\L$ for eigenvalue $\lambda_{k}$, i.e., $\L(\r_{k}) = \lambda_{k} \r_{k}$ and 
$\L^{\dagger}(\l_{k}) = \lambda_{k} \l_{k}$.  In principle $\r_{k} \neq \l_{k}^{\dagger}$ since in general $\L \neq \L^{\dagger}$.
Left and right eigenmatrices form a complete basis, which we normalise as $\Tr (\l_{k} \r_{k'}) = \delta_{k,k'}$. We assume there are no Jordan blocks in the part of the spectrum relevant for our analysis; see e.g.\ Ref.\ \cite{Gaveau2006}.
}. 
The real parts of eigenvalues $\{ \lambda_{k>1} \}$ give the relaxation rates of all the modes of the system dynamics.
In particular, the second eigenvalue $\lambda_{2}$ determines the {\em spectral gap}, whose inverse is related 
to the longest timescale $\tau$ of the relaxation of the system to the stationary state, i.e., $\left\lVert \rt - \rss \right\rVert \sim e^{-t/\tau}$ with $\tau \sim 1/|\Re(\lambda_{2})|$ 
(where $\left\lVert A \right\rVert := \Tr \sqrt{A^{\dagger}A}$).

\begin{figure}[ht!]
\begin{center}
\includegraphics[width=0.8\columnwidth]{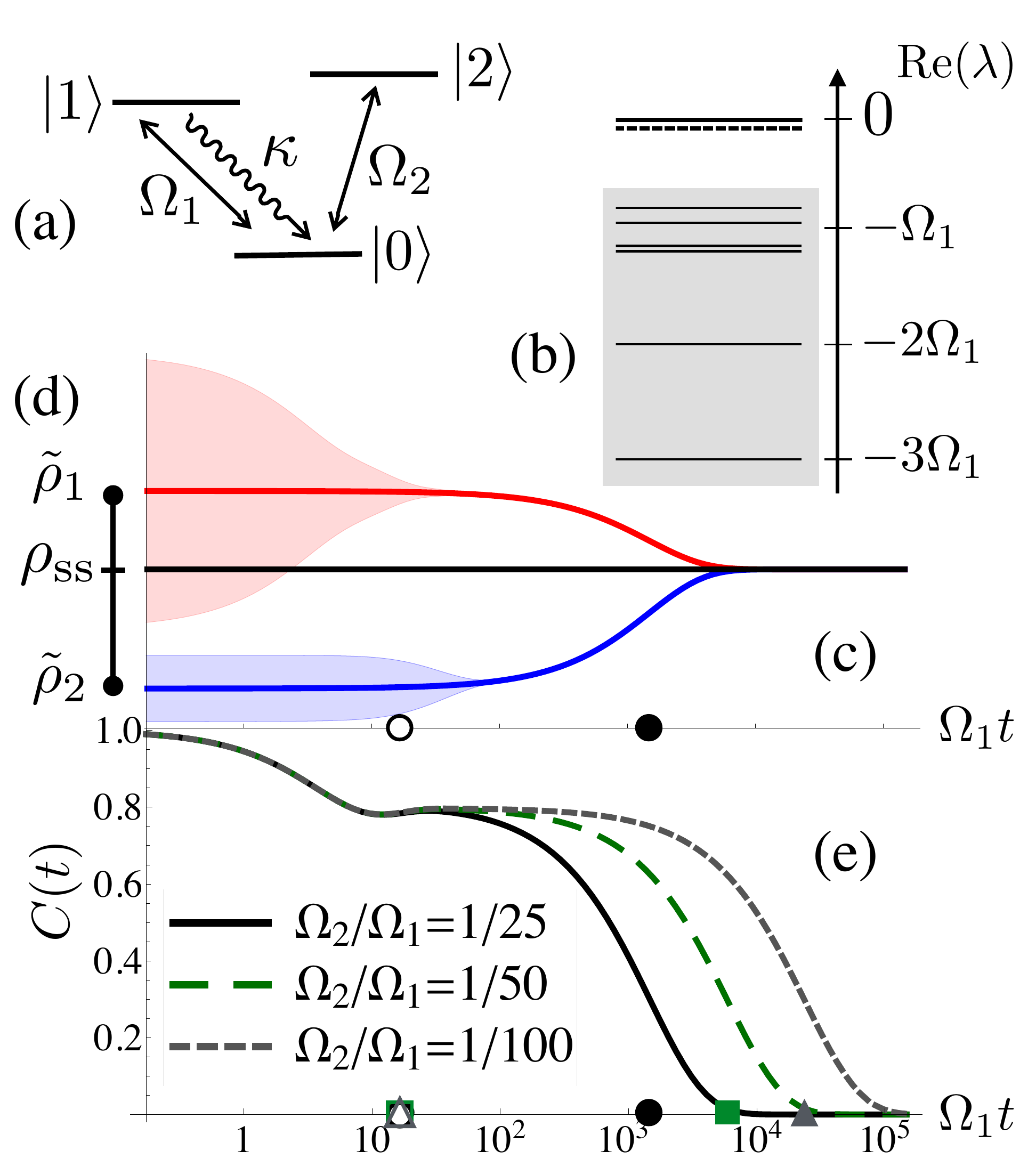}
\caption{
{\bf Example of metastability in a 3-level system:}
(a) Level scheme and transitions. (b) Spectrum of $\L$ showing separation of timescales between $(\lambda_{1},\lambda_{2})$ (full and dashed) and $\{ \lambda_{k>2} \}$ (shaded), for the case $\kappa = 4 \Omega_{1}$, $\Omega_{2} = \Omega_{1}/10$. (c) Illustration of 
the distance of the state, $\rho(t)$, to the MM.  We consider $\rt$ starting from pure states corresponding to the eigenvectors of $\l_{2}$ with maximal (top/red) and minimal eigenvalues (bottom/blue) $c_{2}^{\rm max}$ and $c_{2}^{\rm min}$.  The full curves indicate the nearest state on the MM, $\rms(t)$, to the full state $\rt$.  The shaded region indicates the scale of the ``error'' $\lVert \delta \rt\rVert $ with $\delta \rt := \rt-\rms(t)$.  On times of order $\tau''$ (open circle)
the state $\rho(t)$ relaxes to the MM (in this case to either of the eMS, $\tr_{1,2}$), as seen by the shaded region decreasing to zero.  On times of order $\tau$ (filled circle) there is an eventual relaxation to the stationary state $\rss$ (central/black line). Since $m=2$, in this case $\tau'=\tau$.
(d) The MM is a one-dimensional simplex. 
(e) Normalised autocorrelation, $C(t)$, of the observable  $|1 \rangle \langle 1|-|2 \rangle \langle 2|$, in the stationary state.  For decreasing $\Omega_{2}/\Omega_{1}$ (i.e., decreasing gap), metastability in the regime $\tau''$ (open symbols) to $\tau$ (filled symbols) is increasingly pronounced. }

\vspace*{-4mm}
\label{fig:1}
\end{center}
\end{figure}

Metastability manifests as a long time regime when the system appears stationary, before eventually relaxing to $\rss$.  This occurs when low lying eigenvalues become separated from the rest of the spectrum. Lets assume that this separation occurs between the $m$-th mode and the rest, that is, $|\Re(\lambda_{m})| \ll |\Re(\lambda_{m+1})|$.  We can then write for the time evolution from an initial state $\ri$,
\begin{equation}
\rho(t) = 
e^{t \L} \ri
=
\rss + \sum_{k=2}^{m} e^{t \lambda_{k}} c_{k} \r_k 
+ \left[e^{t\mathcal{L}}\right]_{\I-\P} \ri , 
\label{eq:dyn1}
\end{equation}
where $c_{k}=\Tr(\l_k \ri)$ are coefficients of the initial state decomposition into the eigenbasis of $\L$ \cite{Note2}. 
In \eqref{eq:dyn1} we have introduced the projection $\mathcal{P}$ on the subspace of the first $m$ eigenmatrices, $\mathcal{P} \rho :=\rss \mathrm{Tr}(\rho) + \sum_{k=2}^{m} \r_{k} \mathrm{Tr}(\l_k \rho)$, and $\left[e^{t\mathcal{L}}\right]_{\mathcal{P}}:=\mathcal{P}e^{t\mathcal{L}}\mathcal{P}$.  Expanding the exponentials in the sum, and assuming $\lambda_{1},\ldots,\lambda_{m}$ are real, Eq.\ \eqref{eq:dyn1} can be rewritten as \footnote{
The norm $\left\lVert \cdot \right\rVert$ of a super-operator $\S$,  is the norm induced by the trace norm, $\left\lVert A \right\rVert := \Tr \sqrt{A^{\dagger}A}$, of complex matrices $A$ on which $\S$ acts: $\left\lVert \S \right\rVert := \sup_{\left\lVert A \right\rVert=1} \Tr \left\lVert \S A \right\rVert$.  
 }, 
\begin{eqnarray}
\rho(t) &=& \rss + \sum_{k=2}^{m} c_{k} \r_k 
\nonumber \\
&&
+ \mathcal{O}\left(\left\lVert \left[t \L \right]_{\P}\right\rVert\right)
+ \mathcal{O}\left(\left\lVert \left[e^{t \L} \right]_{\I-\P}\right\rVert\right) .
 \label{eq:dyn2}
\end{eqnarray}
Dynamics will appear stationary for {\em any} initial condition when the last two terms are small.  This defines a range $\tau'' \ll t \ll \tau'$ where metastability occurs. Intuitively the last term can be discarded if $\tau'' \sim 1/|\Re(\lambda_{m+1})|$ and the overlap of the initial state with the suppressed modes is not too large, so that the sum over many modes of small amplitude can be neglected.  Thus, for times $\tau'' \ll t$ the system relaxes into a state in the {\em metastable manifold} (MM).  Apparent stationarity requires
 $\left\lVert \left[t \L \right]_{\P}\right\rVert \ll 1$, which defines the upper limit of the metastable interval: $\tau' \sim 1/|\Re(\lambda_{m})|$ (for $m$ not too large).  
 
More generally, eigenvalues could be complex, appearing in conjugate pairs, $\lambda_{k,1}=\lambda_{k,2}^{*}$, with imaginary parts that cannot be discarded.  Taking this into account, a state $\rms$ in the MM would read in general
\footnote{
For real eigenvalues, $\r_k$ and $\l_k$ can be chosen Hermitian.  Note that while $\r_{1}=\rho_{ss}$, $\r_{k>1}$  are not positive. Complex eigenvalues come in conjugate pairs $\lambda_{k,1}=\lambda_{k,2}^*$ and if so we have $\r_{k,1}=\r_{k,2}^\dagger$, $\l_{k,1}=\l_{k,2}^\dagger$.},
\begin{eqnarray}
\rms &=& \rss + \sum_{k}^{m} c_{k}'(t) \r_k' .
\label{eq:dyn3}
\end{eqnarray}
When $\lambda_k$ is real, we have that $c_k'(t) := c_k$ and $\r_k' := \r_k$.  For conjugate pairs, $\lambda_{k,1}=\lambda_{k,2}^{*}$, we have that $c_{k,1}=c_{k,2}^*$ and $c_{k,1}':=|c_{k,1}| \cos(\omega_k t+\delta_k) $ and $c_{k,2}':=|c_{k,2}| \sin(\omega_k t+\delta_k) $, where $\r_{k,1}':=\r_{k,1}+\r_{k,2}$ and $\r_{k,2}':=i (\r_{k,1}-\r_{k,2})$, with $\delta_k:=\arg (c_k)$, $\omega_k:= \Im{(\lambda_k)}$.   In Eq.\ \eqref{eq:dyn3} we have discarded the second line of Eq.\ \eqref{eq:dyn2}, which leads $\rms$ to be approximately positive with its negative part bounded by the corrections to the invariance of the MM in Eq.\ \eqref{eq:dyn2}. 
The remaining time dependence in Eq.\ \eqref{eq:dyn3} constitutes rotations within the MM that leave the MM invariant, which necessarily correspond to non-dissipative evolution for $\tau'' \ll t \ll \tau'$, which we also discuss below.

Beyond the {\em metastable regime}, $t \gtrsim \tau'$, dynamics will correspond to motion in the MM towards the true stationary state, which is reached at times $t \gg \tau$.  This effective dimensional reduction due to a separation of timescales is a key result of this paper.

\smallskip

\noindent
{\bf \em Geometrical description of quantum metastability.} 
The MM can be described geometrically by generalising the classical method of Refs.\ \cite{Gaveau1996,Bovier2002,Gaveau2006,Nicholson2013,Kurchan2009}.  In the metastable regime the system state is well approximated by a linear combination of the $m$ low-lying modes, see Eqs.\ \eqref{eq:dyn3}. A metastable state is determined by a vector $(c_{2}',\ldots,c_{m}')$ in ${\mathbb R}^{m-1}$. We thus refer to the MM as being $(m-1)$-dimensional, but note that each point on this manifold represents a $D^{2}$ density matrix $\rms$, where $D=\dim(\H)$ is the dimension of the Hilbert space $\H$ of the system. Furthermore, the MM is a convex set as it is a linearly transformed convex  set of initial states $\ri$.  

Let us first consider the case of $m=2$. Due to the convexity of the MM, any metastable state is a mixture of {\em extreme metastable states} (eMS).  In this case they are just two, $\tr_1$ and $\tr_{2}$, obtained from  
\begin{equation}
\tr_{1} = \rss + c_{2}^{\rm max} \r_2 \; , \;\;\; 
\tr_{2} = \rss + c_{2}^{\rm min} \r_2 ,
 \label{eq:ems2}
\end{equation}
where $c_{2}^{\rm max}$, $c_{2}^{\rm min}$ are the maximal and minimal eigenvalues of $\l_{2}$ \cite{Note2}.  Note that $\tr_{1,2}$ are (approximately) positive despite $\r_2$ being non-positive.  From Eq.\  \eqref{eq:dyn2} it follows, up to corrections, that $\rho(t) = p_{1} \tr_{1} + p_{2} \tr_{2}$ with probabilities 
$p_{1,2} = \Tr( \tp_{1,2} \ri )$ where 
\begin{equation}
\tp_{1} = \left( \l_{2} -c_{2}^{\rm min} I \right) / {\Delta c_{2}}\; , \; \;
\tp_{2} = \left( -\l_{2} +c_{2}^{\rm max} I \right) / {\Delta c_{2}} , 
\nonumber
\end{equation}
and $\Delta c_{2} := c_{2}^{\rm max}-c_{2}^{\rm min}$. Note that the observables $\tp_{1,2}$ satisfy $\tp_{1,2} \geq 0$ and $\tp_{1}+\tp_{2}=I$.
This leads to $\tr_{1}$ and $\tr_{2}$ being (approximately) disjoint~\footnote{
See Supplemental Material for derivations of (i) case $m=2$: approximate disjointness of two EMSs and the effective classical dynamics. (ii) class A systems: the structure of the metastable manifold from Eq.\ (\ref{eq:conj}), the metastable regime and the effective dynamics. (iii) class B systems: a comment on the conjecture.}.

\smallskip

\noindent
{\em Example I: 3-level system.} Consider the 3-level system of Fig.\ 1(a), with Hamiltonian $H = \Omega_1 \left( | 1 \rangle \langle 0 | + | 0 \rangle \langle 1 | \right) + \Omega_2 \left( | 2 \rangle \langle 0 | + | 0 \rangle \langle 2 | \right) $ and jump operator $J= \sqrt{\kappa} | 0 \rangle \langle 1 |$.  When $\Omega_{2} \ll \Omega_{1}$, dynamics can be ``shelved'' for long times in $|2\rangle$, giving rise to intermittency in quantum jumps \cite{Plenio1998}, which can be seen as coexistence of ``active'' and ``inactive'' dynamical phases \cite{Garrahan2010}. 
Figure \ref{fig:1}(b) shows the spectrum of $\L$: the gap is small for $\Omega_{2} \ll \Omega_{1}$, the two leading eigenvalues detach from the rest (i.e., $m=2$), and the dynamics is metastable.  Figure \ref{fig:1}(c) illustrates the trace distance of the state $\rt$ to the MM starting from $\ri \neq \rss$: an initial decay on times of order of $\tau''$ to the nearest point on the MM (in this case to an eMS) is followed by decay to $\rss$ on times of order $\tau'=\tau$ (since $m=2$).  The MM for this $m=2$ case is a one-dimensional simplex (i.e., a convex set whose interior points uniquely represent probability distributions on the vertices), see Fig.\ 1(d). 

\smallskip

For $m>2$ the convex set MM of possible coefficients can have more than $m$ extreme points. For classical dynamics it has been proven that this set is well approximated by a simplex \cite{Gaveau2006}, whose vertices correspond to $m$ disjoint eMS and its barycentric coordinates to the probabilities of a metastable state decomposed as a mixture of the eMS, cf.\ Fig.\ 1(d). For quantum dynamics and $m>3$, we expect the structure of the MM to be richer than just a simplex.  As we describe below, the MM can in general also include decoherent free subspaces (DFS) \cite{Zanardi1997d,Zanardi1997e,Lidar1998c} and noiseless subsystems (NSS) \cite{Knill2000b,Zanardi2001b} which are protected from dissipation in the metastable regime, as the next example shows. 

\smallskip

\begin{figure*}[ht!]
\begin{center}
\includegraphics[width=\textwidth]{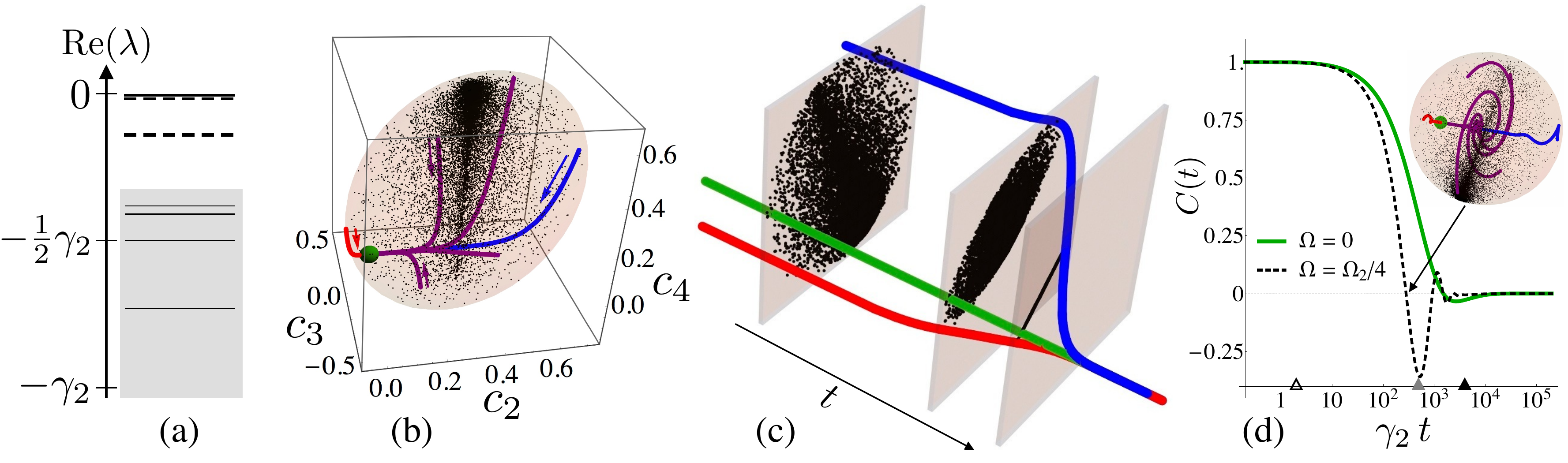}
\caption{
{\bf Example of a coherent metastable manifold:}
(a) Spectrum of $\L$ for Example II (at $\gamma_1=4\gamma_2$, $\Omega_1=2\Omega_2=\gamma_2/50$).
The first four eigenvalues ($m=4$) split from the rest (shaded) and define the MM.  Note the further splitting between $(\lambda_1,\lambda_2)$ and $(\lambda_3,\lambda_4)$ (which are almost degenerate).  (b) The MM is a qubit. Dots represent the metastable states reached from random initial pure states.  They map out under Eq.\ \eqref{eq:dyn3} an affinely transformed Bloch ball (shaded).  The large dot (green) is $\rss$; curves indicate paths in the MM taken by the states evolving from the extreme eigenvectors of $\l_{2}$ (red and blue), $\l_{3}$ and $\l_{4}$ (purple) towards $\rss$.
(c) Time evolution in the MM (affinely transformed to a Bloch ball---planes are projections in the direction of the eigenbasis of $\rss$ and another orthogonal direction): the MM contracts towards a one-dimensional simplex before relaxing eventually to $\rss$, due to the splitting of between the first two eigenvalues and the next two, see panel (a).  
(d) Normalised auto-correlation $C(t)$ for the observable $\sigma_1^z-\sigma_2^z$ (green/solid).  Same for the case where there is an extra perturbing Hamiltonian $\Delta H =\Omega \sigma_1^x \otimes \sigma_2^x$ which induces a rotation in the MM, manifesting in oscillations in $C(t)$ in the metastable regime (black/dashed).  This realises in a metastable system the proposal of \cite{Zanardi2014c,Zanardi2015} for implementing operations in a DFS.
}
\label{fig:2}
\end{center}
\end{figure*}

\noindent
{\em Example II: Collective dissipation and a metastable DFS.} Consider a two-qubit system with Hamiltonian $H = \Omega_{1} \sigma_{1}^{x} + \Omega_{2} \sigma_{2}^{x}$, and a collective jump operator $J= \sqrt{\gamma_1} n_1 \sigma_2^- + \sqrt{\gamma_2} (1-n_{1}) \sigma_2^+$.  When $\Omega_{1,2} \ll \gamma_{1,2}$ there is a small gap and the four leading eigenvalues of $\L$ detach from the rest, Fig.\ \ref{fig:2}(a). 
This is related to the fact that any superposition of $|01\rangle$ and $|10\rangle$ is annihilated by $J$.   Fig.\ \ref{fig:2}(b) maps out the MM by randomly sampling all (pure) initial states  $\ri$ from $\H$ and obtaining their corresponding metastable state via Eq.\ \eqref{eq:dyn3}: the MM is an affinely transformed Bloch ball corresponding to a DFS qubit within the metastable regime $\tau'' \ll t \ll \tau'$.  It important to note: (i) this coherent structure is not the consequence of a symmetry, as for $\gamma_1 \neq \gamma_2$ the system dynamics neither has a U(2) nor an up-down nor a permutation symmetry, cf.~\cite{Albert2014}; (ii) the smallest $m$ for which we can obtain a DFS is $m=4$, as in this case. 

\smallskip

\noindent
{\bf \em Structure of metastable manifold.} 
We aim to find the general structure of the MM for two classes of systems for which $\L$ has a small gap: (A) finite systems where the gap closes at some limiting values of the parameters in $\L$ (such as $\Omega_2 \to 0$ in Example I, and $\Omega_{1,2} \to 0$ in Example II); (B) scalable systems of size $N$ where the gap closes only in the thermodynamic limit $N \to \infty$ (such as the dissipative Ising model of Ref.\ \cite{Ates2012}).  

For class A we prove via non-Hermitian 
degenerate perturbation theory~\cite{Note5} that the structure of a metastable state $\rms \in \text{MM}$ is given by the following block structure,
\begin{equation}
\rms 
= 
\sum_{l=1}^{m'} 
p_l \tr_l \otimes \omega_l 
+
\text{corrections}
, 
\label{eq:conj}
\end{equation}
with $\H$ being the orthogonal sum $\H = \bigoplus_{l} \H_l \otimes \K_l$, where $\tr_l$ are fixed states on $\H_l$ (cf.\ eMS above), $\omega_l$ are arbitrary states on $\K_l$, and $p_{l}$ are probabilities. Up to the corrections, this is a general structure of a manifold of stationary states of open quantum Markovian dynamics~\cite{Baumgartner2008}. The metastable regime is given by $\tau_0\ll t\ll s^{-2}\tau_0,$ where $\tau_0$ is the relaxation time for the unperturbed dynamics and $s$ is the scale of the perturbation~\cite{Note5}. The corrections in Eq.~\eqref{eq:conj} are of the order of the corrections to the invariance of the MM during the metastable time regime, cf.  Eq.~\eqref{eq:dyn2}. The $(m-1)$ coefficients $(c_{2}',\ldots,c_{m}')$ that determine $\rms$, see Eq.\ \eqref{eq:dyn3}, correspond approximately to an affine transformation of the $m$ entries of $p_{l} \omega_{l}$ ($l=1,\ldots,m'$) in Eq.\ \eqref{eq:conj} with $\sum_{l=1}^{m'} p_l=1$~\cite{Note5}. Therefore, the MM approximately represents the degrees of freedom of the classical-quantum space in Eq.\ \eqref{eq:conj}.

For class B we {\em conjecture} that the coefficients representing the MM converge to degrees of freedom of a classical-quantum space as in Eq.\ \eqref{eq:conj}, when the separation in the spectrum becomes more and more pronounced as $N\rightarrow\infty$. Note that the dimensionality of the MM does not change with $N$ and thus the convergence is well defined. 
This general conjecture is based on the {\em necessary condition} that the low-lying spectrum of $\L$ features only trivial Jordan blocks
\footnote{Non-trivial Jordan blocks would lead to an unbounded norm of $\rho(t)$ in the limit of $\text{gap} \to 0$~\cite{Wolf2012}. More precisely, Jordan blocks may be considered as long as they do not contribute significantly, so that Eq.\ \eqref{eq:dyn2} holds true with appropriately redefined corrections being small. 
This condition aims to exclude systems where a small gap is
``accidental'' in the sense it is unrelated to it vanishing in some
limit. 
}.
Note that a conjecture of the $\rms$  structure being approximately that of stationary states, cf. Eq.\ \eqref{eq:conj}, is a stronger claim.  A proof of the former conjecture for class B appears challenging at this moment, see comment in \cite{Note5}. 

The blocks in Eq.\ \eqref{eq:conj} can be of three kinds: (i) When $\dim(\K_l)=1$, the $l$-th block is a disjoint eMS. This is the case in Example I, where there are two eMS, $\tr_{1,2}$, with metastable states being mixtures of them.  
For classical systems the MM is always approximately a simplex of $m$ disjoint eMS \cite{Gaveau2006} with probabilities representing \emph{classical degrees of freedom}.  (ii) When $\dim(\K_l)>1$ and $\dim(\H_l)=1$, $\K_l$ is a decoherence free subspace (DFS) protected from the noise.  This is the case in Example II where the MM is a qubit.  (iii) 
When $\dim(\K_l)>1$ and $\dim(\H_l)>1$, $\K_{l}$ is also protected from noise and termed a noisless subsystem (NSS). 
The structures (ii) and (iii) correspond to \emph{quantum degrees of freedom} ($\omega_l$) and do not appear in the case of classical dynamics \cite{Gaveau2006}.
In general the number of blocks in Eq.\ \eqref{eq:conj} is $m' \leq m$, with equality occurring only when there are no DFS or NSS.

\smallskip

\noindent
{\bf \em Effective motion in the metastable manifold.}
In the metastable regime, $\tau'' \ll t \ll \tau'$, metastable states appear stationary, or perhaps rotate within the MM.  
This latter case corresponds to either: (i) coherent motion in the DFS/NSS where the matrices $\omega_l$  
of Eq.\ \eqref{eq:conj} evolve unitarily in time; or (ii) classical rotations with a frequency which is limited by the dimensionality of MM
\footnote{This latter classical motion was not considered in \cite{Gaveau2006}, as the low-lying eigenvalues of $\L$ were assumed all real.  
One can show \cite{Macieszczak2015} that the set of $(c_{2}',\ldots,c_{m}')$ is again a simplex.  Points within the simplex evolve within it.  Due to exponential shrinking, this evolution can allow for rotations, but with limited frequencies which decrease with decreasing gap.}. 
For class A systems only case (i) is possible~\cite{Note5,Zanardi2014c,Zanardi2015}.

For longer times, $t \gtrsim \tau'$, the MM contracts exponentially towards $\rss$.  This is illustrated in Fig.\ \ref{fig:2}(c) for Example II. This low dimensional evolution in the MM is well described by an effective generator $\mathcal{L}_{\mathrm{eff}}:=\left[\L\right]_{\mathcal{P}}$, which can be considered as the generator of the dynamics averaged over intervals $\tau''$.  If the MM is approximately a simplex (i.e., containing no DFS or NSS) the motion generated by $\mathcal{L}_{\mathrm{eff}}$ is that of classical transitions between {\em macrostates} described by the eMS (see \cite{Note5} for $m=2$ and \cite{Macieszczak2015} for the general case).  
For class A when the MM contains coherent subsystems/subspaces, the motion preserves the structure of Eq.\ \eqref{eq:conj} and can be shown to be trace-preserving and approximately completely-positive~\cite{Note5,Zanardi2016,Zi2013}. Note that decoupling of (slower) classical dynamics from (faster) quantum evolution in the MM requires further separation in low-lying eigenvalues of $\mathcal{L}$. This is illustrated in Fig.\ \ref{fig:2}(c) for Example II. 

In practice, metastability can be accessed through the connected auto-correlation~\cite{Sciolla2015}
of the measurement $\M$ of a system observable, even in the stationary state, $C(t) := \Tr(\M e^{{t \L}} \M \rss) - \Tr(\M \rss)^{2}$ 
\footnote{
For an observable ${M = \sum_{i=1}^{D} m_i | m_{i} \rangle \langle m_{i}|}$, we have that ${\M(\rho) = \sum_{i=1}^{D} m_i \langle m_{i}|\rho|m_{i}\rangle  |m_{i}\rangle \langle m_{i}|}$.};
see Figs.\ \ref{fig:1}(e), \ref{fig:2}(d).  The first measurement $\M$ perturbs $\rss$, and the  state conditioned on the result partially relaxes towards the MM for $t \lesssim \tau''$. In the metastable regime correlations will persist as the different blocks in \eqref{eq:conj} do not communicate, and for the case where all low-lying eigenvalues are real, $C(t) \approx \Tr(\M \P \M \rss) - \Tr(\M \rss)^{2}$.  When low-lying eigenvalues are complex, oscillations of $C(t)$ can occur in the metastable regime, as in Fig.\ \ref{fig:2}(d). 
When $t\gtrsim \tau'$, dynamics begins to relax back towards $\rss$,  erasing all information about the initial result, $C(t) \approx 0$, for  $t \gg \tau$.

\smallskip

\noindent
{\bf \em Outlook.} The next steps in the development of the theory of quantum metastability presented here include:

(i) For many-body systems, where direct diagonalisation of $\L$ is impractical, it should be possible to use dynamical large-deviation methods \cite{Touchette2009} to identify dynamically the different blocks in Eq.\ \eqref{eq:conj} by biasing ensembles of quantum trajectories \cite{Garrahan2010}. This approach could be implemented numerically 
by generalising classical path sampling \cite{Hedges2009} and/or cloning techniques \cite{Giardina2011}.

(ii) In order to reveal the structure of the MM, one needs to find a general computational scheme that can identify the basis in which metastable states look explicitly as in Eq.\ \eqref{eq:conj}.  Such a method would be useful to uncover DFS and NSS more generally.  Also, it would be interesting to consider more broadly DFS that do no arise as a consequence of symmetry, cf.\ Example II above. 

(iii) We have considered here metastability in the case of Markovian dynamics generated by a Lindbladian $\L$.  Metastability occurs also when dynamics is non-Markovian, see e.g.\ \cite{Merkli2015}.  It should be possible to generalise the method introduced above to the non-Markovian case of a time-dependent generator $\L(t)$. 

(iv) A significant challenge is to extend the ideas presented here to study metastability in {\em closed} quantum systems.  This would be relevant to the fundamental problems of thermalisation \cite{Eisert2014} and many-body localisation \cite{Nandkishore2015}.

\smallskip

\begin{acknowledgments}
This work was supported by EPSRC Grant No.\ EP/J009776/1 and ERC Grant Agreement No.\ 335266 (ESCQUMA). K.M. thanks M. Idel for discussions.
\end{acknowledgments}

\bibliography{metastability2}

\pagebreak

\onecolumngrid

\newcommand{\hz}{{H^{(0)}}}
\newcommand{\hf}{{ H^{(1)}}}
\renewcommand{\j}{{J_{j}}}
\newcommand{\jz}{{J_{j}^{(0)}}}
\newcommand{\jf}{{ J_{j}^{(1)}}}

\newcommand{\rz}{{\r_k^{(0)}}}
\newcommand{\lz}{{\l_k^{(0)}}}

\newcommand{\Lf}{{\L^{(1)}}}
\newcommand{\Ls}{{\L^{(2)}}}
\newcommand{\tLf}{{\widetilde{\L}^{(1)}}}
\newcommand{\tLs}{{\widetilde{\L}^{(2)}}}
\renewcommand{\Q}{{\mathcal Q}}
\newcommand{\Z}{{\mathcal Z}}
\newcommand{\T}{{\mathcal T}}
\renewcommand{\O}{{\mathcal O}}

\newcommand{\Tz}{{\T^{(0)}}}
\newcommand{\Tf}{{\T^{(1)}}}
\newcommand{\Ts}{{\T^{(2)}}}

\newcommand{\Pz}{{\P}}
\newcommand{\Pf}{{\P^{(1)}}}
\newcommand{\Ps}{{\P^{(2)}}}

\newcommand{\Lz}{{\L}}

\newcommand{\tS}{{\widetilde{S}}}
\newcommand{\tLt}{{\widetilde{\L}^{(3)}}}
\newcommand{\LS}{{\L(s)}}
\newcommand{\PS}{{\P(s)}}
\newcommand{\tSl}{{\widetilde{\S}_l}}
\newcommand{\Pl}{{\P_l}}
\newcommand{\PlS}{{\Pl(s)}}
\newcommand{\Plj}{{\P_{l,j}}}
\newcommand{\PljS}{{\Plj(s)}}

\newcommand{\tLS}{{\widetilde\L(s)}}

\setlength{\parskip}{2mm}
\setlength{\parindent}{0mm}

\section*{Supplemental Material}

\subsection*{Metastability for two low-lying eigenmodes}

Here we consider the case of $m=2$ low-lying eigenvalues in the master operator $\L$, see Eqs.~(1-4) in the main text. \\

Since the metastable manifold (MM) is convex and 1-dimensional, it is simply an interval and thus a simplex.  Hence, any metastable state is a mixture of  extreme metastable states (eMSs), in this case two: $\tr_{1}$, $\tr_{2}$. As a metastable state, $\rms=\rss+c_2\r_2$, is determined by the coefficient $c_2=\Tr(\l_2\ri)$, the eMSs correspond to the extreme values of $c_2$ given by the maximum $c_{2}^{\max}$ and minimum $c_{2}^{\min}$ eigenvalue of $\l_2$, see Eq. (5) in the main text. Furthermore, note that $\tr_{1}$, $\tr_{2}$ are the metastable states for the pure initial state given by the $\l_2$ eigenvectors corresponding to $c_{2}^{\rm max}$ and $c_{2}^{\rm min}$, respectively. As $\tr_{1}, \tr_{2}$ are then given by the truncated evolution equation (3) of the main text, we have $\mathrm{Tr}(\tr_{1,2})=1$ and $\tr_{1}, \tr_{2}$ are approximately positive with the corrections bounded by the corrections to stationarity in the metastable regime.   

The decomposition $\rms = p_{1} \tr_{1} + p_{2} \tr_{2}$ of a metastable state into the eMSs is given by the observables 
$\tp_{1}$, $\tp_{2}$ (for definition see the main text below Eq. (5)) which determine the probabilities as  $p_{1,2} = \Tr( \tp_{1,2} \,\ri )$. We note that  the definition of $\tilde{\rho}_{1,2}$ and $\tr_{1,2}$ insures that 
$\mathrm{Tr}(\tp_i\tr_j)=\delta_{ij}$ for $i,j=1,2$, and $\tp_{1,2} \geq 0$ and $\tp_{1}+\tp_{2}=\mathds{1}_\H$, i.e., $\{\tp_{k}\}_{k=1}^2$  constitute a POVM.


\noindent
{\bf Approximate disjointness of two eMS.} Below we prove that the extreme metastable states are approximately disjoint. More precisely, we show that there is a division of the system Hilbert space $\mathcal{H}=\mathcal{H}_{1}\oplus\mathcal{H}_2$ so that $\Tr \left(\mathds{1}_{\mathcal{H}_{1,2}} \tr_{1,2}\right)\geq 1-\mathcal{O}(C)$, where $C$ are the corrections to the stationarity in the metastable regime, cf. Eq. (3) in the main text.

\emph{Proof.} Note that the stationary state $\rss$ is a mixture of the two eMS, $\rss=p_1^{ss}\,\tr_{1}+p_2^{ss}\, \tr_{2}$, where $p_1^{ss}=-c_2^{\min}/\Delta c_2$ and $p_2^{ss}=c_2^{\max}/\Delta c_2$. We define the orthogonal subspaces $\mathcal{H}_{1}$ and $\mathcal{H}_2$ as follows,
\begin{eqnarray}
\mathcal{H}_1&=&\mathrm{span}\, \{ |\psi_k\rangle, k=1,..,D:\, \langle\psi_k|\tp_1 |\psi_k\rangle\geq p_1^{ss}\}, \\
\mathcal{H}_2&=&\mathrm{span}\, \{ |\psi_k\rangle, k=1,..,D:\, \langle\psi_k|\tp_2 |\psi_k\rangle > p_2^{ss}\},
\end{eqnarray}
where $\{|\psi_k\rangle\}_{k=1}^D$ is the orthonormal eigenbasis of $\l_2$, which is also the eigenbasis of both $\tp_1$ and $\tp_2$. From $\tp_2=\mathds{1}-\tp_1$, we have $\mathcal{H}=\mathcal{H}_{1}\oplus\mathcal{H}_2$.  Let $|\psi_1 \rangle$ and $|\psi_2 \rangle$ denote the eigenvectors of $\l_2$ corresponding to the extreme eigenvalues $c_{2}^{\max}$ and $c_{2}^{\min}$. Let $\rho_1(t)$, $\rho_2(t)$  further be the system state at time $t$ for the initial state chosen as  $|\psi_1\rangle$,$|\psi_2\rangle$. From the orthogonality of the  eigenmatrices of $\L$ (also in the case of Jordan blocks in $\I-\P$), it follows that
\begin{equation}
\Tr \left(\tp_1\rho_2(t)\right)=\Tr \left(\tp_2\rho_1(t)\right)=p_1^{ss}\,( e^{t\lambda_2}-1). \label{eq:ortho1}
\end{equation}
From positivity of $\rho_1(t)$ and the fact that $\mathds{1}_{\mathcal{H}_1}$ is diagonal in the eigenbasis of $\tp_1$, we also have
\begin{equation}
\Tr \left(\tp_1\rho_2(t)\right)\geq \Tr \left( \mathds{1}_{\mathcal{H}_1}\tp_1 \rho_2(t)\right)\geq p_1^{ss}\,\Tr \left( \mathds{1}_{\mathcal{H}_1}\rho_2(t)\right),
\end{equation}
Together with Eq.~\eqref{eq:ortho1} it follows that
\begin{equation}
\Tr \left( \mathds{1}_{\mathcal{H}_1}\tr_2\right)\leq \Tr \left( \mathds{1}_{\mathcal{H}_1}\rho_2(t)\right)+\mathcal{O(C)}\leq ( e^{t\lambda_2}-1)+\mathcal{O(C)}=\mathcal{O(C)},
\end{equation}
where $C$ are the corrections to the stationarity in the metastable regime, cf. Eq. (3). Analogously,   $\Tr \left( \mathds{1}_{\mathcal{H}_2}\tr_1\right)\leq \mathcal{O(C)}$, which ends the proof. Let us note that this argument is analogous to the case of $m=2$ in classical systems~\cite{g2}. $\blacksquare$

\noindent
{\bf Effective classical dynamics in the metastable manifold.} Here we consider the linear operator $\Leff:=[\L]_{\P}$ which governs the dynamics for times $t\gtrsim\tau'$. Note that in this case, $m=2$, we have simply $\tau=\tau'=(-\Re\lambda_2)^{-1}$. 

Note that by the construction, the operator  $\Leff$ transforms the MM into itself. As  for $m=2$ the MM is a simplex, $\Leff$ generates a \emph{positive} and \emph{probability preserving} evolution of the probabilities $(p_1(t),p_2(t))$. This implies that $\Leff$ is a generator of \emph{classical stochastic dynamics}.  Indeed, in the basis of extreme metastable states, $\tr_1$, $\tr_2$, we have 
\begin{equation}
\frac{\mathrm{d}}{\mathrm{d}t}\left( \begin{array}{c} p_1(t) \\ p_2(t) \end{array} \right)=\Leff \left( \begin{array}{c} p_1(t) \\ p_2(t) \end{array} \right)=\frac{1}{\Delta c_2}\left( \begin{array}{cc} c_{2}^{\max}\, \lambda_2  & c_{2}^{\min}\, \lambda_2 \\
-c_{2}^{\max}\,\lambda_2 & - c_{2}^{\min}\,\lambda_2 \end{array} \right)
\left( \begin{array}{c} p_1(t) \\ p_2(t) \end{array} \right),
\label{eq:Leff2}
\end{equation}
where $\Delta c_2= c_{2}^{\max}-c_{2}^{\min}$. We have that $\lambda_2<0$ and  $c_{2}^{\min}\leq 0$ due to $\Tr(\l_2\rho_{ss})=0$. Therefore it follows that  $\Leff$ indeed obeys the generator characteristics:  the diagonal terms are negative, the off-diagonal terms are positive and the sum of entries in each column is 0. The corresponding dynamics is thus given by 
\begin{equation}
\left( \begin{array}{c} p_1(t)\\ p_2(t) \end{array} \right)= e^{t\Leff} \left( \begin{array}{c} p_1\\ p_2 \end{array} \right)=\frac{1}{\Delta c_2}\left( \begin{array}{cc}c_{2}^{\max} \,e^{t\lambda_2} - c_{2}^{\min} & - c_{2}^{\min}\, (1-e^{t\lambda_2}) \\
c_{2}^{\max}\, (1-e^{t\lambda_2} ) & c_{2}^{\max} - c_{2}^{\min}\,e^{t\lambda_2} \end{array} \right)
\left( \begin{array}{c} p_1 \\ p_2 \end{array} \right),
\label{eq:dyn4}
\end{equation}
and for $t\rightarrow\infty$ we obtain the probabilities corresponding to the stationary state, $(-c_{2}^{\min}\tr_1+c_{2}^{\max}\tr_2)/\Delta c_2 = \rho_{ss}$. The dynamics in~\eqref{eq:dyn4} approximates the system dynamics with the corrections being bounded by the corrections to stationarity in the metastable regime (cf. Eq.~(2) in the main text),
\begin{equation}
\rho(t)=e^{t\mathcal{L}}\ri= p_1(t)\,\tr_1+p_2(t)\,\tr_2 +\mathcal{O}\left(\left\lVert \left[e^{t \mathcal{L}}\right]_{\mathcal{I-P}}\right\rVert \right).
 \label{eq:dyn2}
\end{equation}
Let us finally emphasize that for times $t\gtrsim\tau'$ dynamics takes place between eMSs, $\tr_1$, $\tr_2$, which can be considered as system \emph{macrostates} in analogy to classical thermodynamics. The generator in Eq.~\eqref{eq:Leff2} yields \emph{stochastic trajectories} of transitions between  $\tr_1$, $\tr_2$. Those trajectories correspond to quantum trajectories \emph{coarse-grained in time} over intervals of the order $\mathcal{O}(\tau'')$, similarly as in the example of 3-level atom, see Fig. 1 in the main text, the intermittency in quantum jumps corresponds to conditional system dynamics being restricted to the dark level $|2\rangle$ (``inactive" dynamics) or the subspace spanned by the level $|1\rangle$ and $|0\rangle$ (``active'' dynamics)~\cite{M}.

\subsection*{Characterising the structure of the metastable manifold and effective dynamics for Class A systems}

In this section we discuss metastablity of a finite open quantum system for which the gap closes at some value of parameters in the master equation (see Eq.~(1) in the main text) so that the stationary state is no longer unique.  For  dynamics which are close to the degenerate case, we prove that there is a separation in the spectrum leading to a metastable time regime during which the system's state has the structure given in Eq.~(6) of the main text. Moreover, the effective dynamics in the metastable manifold is trace-preserving and approximately completely positive.  \\


\noindent
{\bf Perturbation theory analysis.} We use the perturbation theory of linear operators (see Chapter 2 of~\cite{K95}) in order to analyse an open quantum system of finite dimension whose Lindblad operator $\L(s)$ is obtained by perturbing a generator $\Lz= \L(0)$ featuring multiple stationary states. We consider $\Lz$ with $m$-fold degeneracy of the stationary state manifold (SSM). In the proof we assume that the dynamics exhibits no rotations in the stationary state manifold, i.e. $\Lz$ has no non-zero imaginary eigenvalues. The case of unitarily rotating SSM can be analysed in a similar fashion \cite{M}. Consequently, there are $m$ right (left) eigenmatrices corresponding to the $0$ eigenvalues, with no non-trivial Jordan blocks due to positive and trace-preserving dynamics~\cite{W}. 
 The asymptotic states of $\Lz$ have the structure given by Eq.~(6) in the main text (without the corrections),  
 see e.g. \cite{B08}. 
 We denote by $\Pz$ the projection on the SSM of $\Lz$, with $\Pz (\cdot)=\sum_{l=1}^{m'} \rho_{l}\otimes \Tr_{\H_l}(\mathds{1}_{\H_l}\otimes\mathds{1}_{\K_l}\,(\cdot)\,)$, so that for the initial state $\ri$, the asymptotic state is given by $p_l=\Tr(\mathds{1}_{\H_l}\otimes\mathds{1}_{\K_l}\,\ri)$ and $\omega_l=\Tr_{\H_l}(\mathds{1}_{\H_l}\otimes\mathds{1}_{\K_l}\,\ri)/p_l$, $l=1,...,m'$.

For simplicity, we consider a linear perturbation of the Hamiltonian $H(s)=H+s\hf$, where $\hf$ is Hermitian, and of the jumps operators are $\j (s)=J_j+s\jf$. The derivations below can be easily generalised to any analytic perturbation of $H$ and $\j$~\cite{K95}. This leads to the following first- and second-order perturbation for the generator 
\begin{eqnarray}
 \LS &=& \Lz+ s\Lf+ s^2\Ls,\quad\mathrm{where}  
\nonumber\\
\Lf &=& -i[\hf,(\cdot)] + \sum_{j} \left(\jf (\cdot) J_j^\dagger - \frac{1}{2}\left\{ J_j^\dagger \jf , (\cdot)\right\} + \mathrm{h.c.}\right)
\nonumber\\
\Ls &=&  \sum_{j} \left(\jf (\cdot) \jf^\dagger - \frac{1}{2}\left\{\jf ^\dagger \jf , (\cdot)\right\} \right).
\label{eq:Lseries}
\end{eqnarray}
We choose the dimensionless scale parameter $s$ so that that $\max(\lVert\Lf\rVert,\lVert\Ls\rVert)=\O(\tau^{-1})$, where $\tau$ is the relaxation time for $\Lz$ dynamics (see below Eqs.~\eqref{eq:Ppert}-\eqref{eq:Lpert} for the precise definition of $s$).

From the perturbation theory of linear operators~\cite{K95}, the eigenvalues of the perturbed operator $\LS$  are continuous with respect to $s$. Furthermore, if $\lambda$ is an eigenvalue  of $\Lz$ with algebraic multiplicity $m$, then for $s$ small enough $m$ eigenvalues of $\LS$ will cluster  around the unperturbed eigenvalue $\lambda$. Those eigenvalues are referred to as the $\lambda$-group.  In general the individual eigenvalues in the $\lambda$-group are not analytic in $s$, but correspond to branches of analytic functions. Moreover, the corresponding eigenmatrices may feature poles.  However, the projection onto the subspace spanned by the $\lambda$-group eigenmatrices is analytic and it follows that the restriction of $\LS$ to this subspace is analytic as well. When $m=1$, the eigenvalue $\lambda(s)$ and the projection on the corresponding eigenmatrix is analytic.  
  
In particular, for $s$ small enough, the first $m$ eigenvalues of $\LS$ belong to $0$-group clustering around $0$ and the separation to the $(m+1)$-th eigenvalue is maintained. Let $\PS $ be the analytic projection on the $0$-group, (which is denoted by $\P$ in the main text for a generic system). Then the restricted generator is given by $[\LS]_{\PS}:=\PS \LS \PS$. Since there are no non-trivial Jordan block associated with the $0$-eigenvalue of $\Lz$~\cite{W}, we have~\cite{K95}
\begin{eqnarray}
\PS &=&\Pz + s\left(-\S \Lf \Pz -\Pz \Lf \S   \right)+ \mathcal{O}(s^2)=:   \Pz + s \Pf+ \mathcal{O} (s^2) , \label{eq:Ppert} 
\\
\left[{\LS}\right]_{\PS}
&=& 
\label{eq:Lpert}
s\,[\Lf]_{\Pz}+s^2\left([\Ls]_{\Pz}-\Pz\Lf \S\Lf\Pz - \S\Lf [\Lf]_{\Pz} - [\Lf]_{\Pz} \Lf \S \right)+\mathcal{O} (s^3(\lVert\Lf\rVert+\lVert\Ls\rVert)))\\
&=:& s\,\tLf+s^2 \,\tLs + \mathcal{O} (s^3(\lVert\Lf\rVert+\lVert\Ls\rVert)))\nonumber,
\end{eqnarray}
where $\S$ is the reduced resolvent of $\Lz$ at $0$, i.e. $\S \,\Lz=\Lz\,\S=\I-\Pz$ and  $\S\,\Pz=\Pz\,\S=0$. The resolvent $\S$ is related to the relaxation time,   $\lVert \S\rVert=\mathcal{O}(\tau)$. 
We now \emph{define} the scale $s$ of the perturbation in Eq.~\eqref{eq:Lseries} so that 
$\max(\lVert\Lf\rVert,\lVert\Ls\rVert)=\lVert \S\rVert^{-1}$, and we will make repeated use of this bound below.

\bigskip

\noindent
{\bf Spectrum of $\LS$}. As we show below, from the fact that both $\Lz$ and $\LS$ are completely positive trace-preserving (CPTP) generators, it follows that first $m$-eigenvalues of $\LS$ are not only continuous, but differentiable continuously at least twice, i.e., 
\begin{equation}
\lambda_k(s)=s\lambda_k^{(1)}+s^2\lambda_k^{(2)}+o(s^2(\lVert\Lf\rVert+\lVert\Ls\rVert)), \quad k=1,...m.\label{eq:Lambdapert}
\end{equation}
Moreover, we have that $\Re\lambda_k^{(1)}=0$ and $\Re\lambda_k^{(2)}\leq 0$, so that the spectrum structure of a positive trace-preserving generator is reproduced in the second order of the perturbation theory. This is due to the fact that the first-order correction is an eigenvalue of $[\Lf]_{\Pz}$, which is a unitary generator~\cite{Z1,Z2} and the second-order correction is an eigenvalue of a CPTP generator on the SSM of $\Lz$ (see also~\cite{Z3}).

In the generic case when the degeneracy of the first $m$-eigenvalues is lifted in the second order of the perturbation theory, we further demonstrate that all $\lambda_k(s)$ are actually analytic in $s$ and so are the projections on the corresponding eigenmatrices, $P_{k}(s)\,(\cdot):=\r_k(s)\,\Tr(\l_k(s)(\cdot))$.  Note that in this case, the stationary state of $\LS$ for $s>0$ is necessary unique, as considered in the main text.\\


\vspace{1mm}

\emph{First-order perturbation}. Let $\tLf:=[\Lf]_{\Pz}$. As we show at the end of this section $\tLf$ is a CPTP generator on the SSM of $\Lz$, and thus its eigenvalues have non-positive real parts. From the definition of $\LS$ we see that also $\L(-s)$ is a CPTP generator, but its first-order correction is of the opposite sign. Hence, $\tLf$ eigenvalues must be imaginary and there is no dissipation. Indeed, in~\cite{Z1,Z2} it was shown that the first order yields \emph{unitary dynamics} and the formula for the corresponding Hamiltonian was derived. 

\vspace{1mm}

\emph{Second-order perturbation}. Let $\tLs:=[\Ls]_{\Pz}-\Pz\Lf \S\Lf\Pz - \S\Lf [\Lf]_{\Pz} - [\Lf]_{\Pz} \Lf \S.$ The generator $\tLf$  lifts \emph{partially} the degeneracy of the $m$-eigenvalues. From Eq.~\eqref{eq:Lpert},  analogously as in the Hermitian perturbation theory, in order to further lift the degeneracy the higher-order corrections should be considered separately for each eigenprojection of $\tLf$. This corresponds to the \emph{reduction process}~\cite{K95} in which, instead of $[\LS]_{\PS}$, one equivalently considers the perturbation theory for $s^{-1}[\LS]_{\PS}=\tLf+s\tLs+\mathcal{O} (s^2(\lVert\Lf\rVert+\lVert\Ls\rVert))$ with the unperturbed operator $\tLf$ and an analytic perturbation, cf. Eq. \eqref{eq:Lpert}. The eigenvalues of $s^{-1}[\LS]_{\PS}$ are related to $\lambda_1(s)$, ..., $\lambda_m(s)$ of $\LS$ simply be multiplication by $s^{-1}$. Since the unitary generator $\tLf$ features only trivial Jordan blocks,  for the eigenspace related to its $\lambda^{(1)}_l$ eigenvalue we obtain that (cf. Eqs.~\eqref{eq:Ppert},~\eqref{eq:Lpert})
\begin{eqnarray}
\Pl(s)&=&\Pl\,+\,s\left(-\Pl\tLs \tS_l  -\Pl\Lf \S + ({\rm inv.})\right)+\mathcal{O} (s^2)\nonumber\\
&=&\Pl\,+\,s\left(-\Pl\Ls \tS_l  -\Pl\Lf \S \Lf \tS_l -\Pl\Lf \S +({\rm inv.})\right)+\mathcal{O} (s^2)\label{eq:Ppert2},
 \\
\left[ \LS \right]_{\PlS}&=&s\,\lambda^{(1)}_l\Pl+s^2\,\left[\tLs\right]_{\Pl}+\mathcal{O} (s^3))\nonumber\\ 
&=&s\,\lambda^{(1)}_l\Pl+s^2\left([\Ls]_{\Pl}-\Pl\Lf \S\Lf\Pl \right)+\mathcal{O} (s^3),\quad l=1,...,m''.\label{eq:Lpert2}
\end{eqnarray} 
Above, $\Pl$ denotes the projection on the $\lambda^{(1)}_l$-eigenspace of $\tLf$, so that we have $\sum_{l=1}^{m''} \Pl=\Pz$. 
Also, $\PlS$ is the projection on the $\lambda^{(1)}_l$-group and $\tS_l=\sum_{k=1, k\neq l}^{m''}(\lambda_k^{(1)}-\lambda_l^{(1)})^{-1} \Pl$ is the reduced resolvent for $[\tLf]_{\Pl}$  at $\lambda_{l}^{(1)}\neq0$, restricted to $\Pz$. Finally, (inv.) denotes the terms with the inverted order of operators. 

From Eq.~\eqref{eq:Lpert2} we see that the degeneracy of the $m$ eigenvalues can be further lifted by the operator $[\tLs]_{\Pl}$. Due to the \emph{reduction process} the eigenvalues of $\LS$ from $0$-group are of the form $s\lambda_l^{(1)}+s^2\lambda_{l,j}^{(2)}(s)=s\lambda_l^{(1)}+s^2\lambda_{l,j}^{(2)}+o(s^2)$, where $\lambda_{l,j}^{(2)}$ is an  eigenvalue of  $[\tLs]_{\Pl}$ and $\lambda_{l,j}^{(2)}(s)$ is the corresponding eigenvalue of  $s^{-1}(s^{-1}[\LS]_{\PlS}-\lambda^{(1)}_l\Pl)=[\tLs]_{\Pl}+s\tLt_l+\mathcal{O}(s^2)$ (see Eq.~\eqref{eq:L3tildeL}).  Below we show that $\Re\,\lambda_{l,j}^{(2)}\leq0$, which ends the proof of Eq.~\eqref{eq:Lambdapert}. Moreover, when the eigenvalues of $[\tLs]_{\Pl}$ are non-degenerate, the corresponding perturbed eigenvalues, $\lambda_{l,j}^{(2)}(s)$, are analytic in $s$ and thus the $0$-group eigenvalues of $\LS$ are analytic. Furthermore, the projection $\Plj(s)$ on the eigenmatrix corresponding to $\lambda_{l,j}^{(2)}(s)$ is analytic and since it is also a projection on the eigenvalue from the $0$-group, the projections on the $m$ low-lying eigenvalues of $\LS$ are analytic. 

\vspace{1mm}

We argue now that $\Re\,\lambda_{l,j}^{(2)}\leq0$. We use the fact proven at the end of this section that $[\tLs]_{\Pz}$ is a CPTP generator on the SSM of $\Lz$. The restricted operator $[\tLs]_{\Pl}$ can be related to $[\tLs]_{\Pz}$ as follows,
\begin{equation}
\sum_{l=1}^{m''}\,\left[\tLs\right]_{\Pl}\,=\,\lim_{t\rightarrow\infty}t^{-1}\int_{0}^t \mathrm{d}u\, e^{-u\tLf}\left[\tLs\right]_{\Pz}e^{u\tLf}.\label{eq:Lpert2int}
\end{equation}
Note that $e^{-u\tLf}[\tLs]_{\Pz}\,e^{u\tLf}$  is the \emph{interaction picture} for $[\tLs]_{\Pz}$. Hence it is a CPTP generator on the SSM of $\Lz$ and  $\sum_{l=1}^{m''}\,[\tLs]_{\Pl}$ as an integral of CPTP generators is also a CPTP generator on the SSM. Moreover, the eigenvalues of $\sum_{l=1}^{m''}\,[\tLs]_{\Pl}$ obey $\Re\,\lambda_{l,j}^{(2)}\leq 0$, which ends the proof.  Note that Eq.~\eqref{eq:Lpert2int} is  the first-order perturbation theory for weak dissipation, where the fast unitary evolution given by $\tLf$ erases all the contributions of the slow dissipation $s\left[\tLs\right]_{\Pz}$ that would create any coherence with respect to the eigenbasis of the Hamiltonian governing the unitary evolution.  $\blacksquare$\\

\bigskip

\noindent
{\bf Time regime of metastability.} We now discuss how the perturbations in Eq.~\eqref{eq:Lseries}  change the system dynamics. We derive the metastable regime when the system dynamics appears stationary as a consequence of the separation in the spectrum of $\L(s)$ discussed above. Let us consider separately the low-lying modes,  given by the projection $\PS$, and the rest of modes  (cf. Eq. (2) in the main text)
\begin{equation}
e^{ t\LS} =
[e^{ t\LS}]_{\PS}
+
[e^{ t\LS}]_{\I-\PS}.
\label{eq:PIPI}
\end{equation}

{\bf Timescale $\tau'(s)$}. By definition, the dynamics maps the MM defined by $\PS$ into itself. However, in the metastable regime, the system dynamics leaves the MM approximately invariant, in the sense that its image is well approximated by the MM itself. This defines the longer timescale $\tau'(s)$ of the regime (see the main text). 
As the first-order correction $s\tLf$ to $[\LS]_{\PS}$ in Eq.~\eqref{eq:Lpert} corresponds to the unitary dynamics leaving the SSM of $\Lz$ invariant, the timescale $\tau'(s)$ will be related to higher-order corrections in $s$, $[\LS]_{\PS} - s\tLf$, cf. Eq \eqref{eq:Ppert}. Indeed, below we show that the corrections to the invariance of the MM are given by
\begin{eqnarray}
[e^{ t\LS}]_{\PS}&=&e^{ t s\tLf} \Pz +s\left(-\S \Lf e^{ t s\tLf} \Pz-e^{ t s \tLf} \Pz  \Lf \S\right)\,+\,\O(s^2)\,+\label{eq:msregime}\\
&&\,+\,t \,s^2\,  e^{t s\tLf}\,\left(t^{-1}\int_0^t\mathrm{d}u\, e^{-u s\tLf}\,\left[\tLs\right]_{\Pz}\,e^{u s\tLf}\right)\,+\,t\,\O(s^3\,\lVert \tLs\rVert)\,+\, t^2\,\O(s^4 \lVert \tLs\rVert^2).\nonumber
\end{eqnarray}
The first line describes unitary dynamics in the metastable manifold, whereas the second line is the contribution from the dissipative dynamics (in the interaction picture). Therefore, the metastable regime is limited to times $t$ for which  all three terms on the second line are small. Since terms are bounded by $ts^2 \|\tLs \|,\mathcal{O}( ts^3  \|\tLs \|)$, and respectively $\mathcal{O}((ts^2 \|\tLs \|)^2 )$ the condition is satisfied 
if $t\ll\tau'(s)$, where 
\begin{eqnarray}
\tau'(s)&=&\left(s^{-2}  +\,\mathcal{O}(s^{-1}) \right) \left\lVert\tLs\right\rVert^{-1}\,\geq\,(s^{-2}+\,\mathcal{O}(s^{-1}))\,\left(\lVert \Ls\rVert+\lVert \Lf\rVert^2 \lVert\S\rVert\right)^{-1}\nonumber\\
&\geq&(s^{-2}+\,\mathcal{O}(s^{-1}))\,\left(\lVert \Ls\rVert+\lVert \Lf\rVert\right)^{-1}\,=\,s^{-2} \O(\tau)\,+\,\mathcal{O}(s^{-1}\tau),\label{eq:taup}
\end{eqnarray}
Here we used Eq. \eqref{eq:Lpert} and the definition of the scaling $\max(\lVert\Lf\rVert,\lVert\Ls\rVert)=\lVert \mathcal{S}\rVert^{-1}=\O(\tau)^{-1}$ to conclude that $\| \tLs \|  \leq O(\tau)^{-1}$, and the Taylor expansion in the first line. 
Note that for small $s$ the leading term of the metastable range is 
$\mathcal{O}(\tau/s^2)$.

\vspace{1mm}

\emph{Derivation of Eq. \eqref{eq:msregime}}. The proof below is analogous to the results of the appendix in~\cite{Z2}. Note that for times $t\ll\tau' (s)$ the unitary contribution to the dynamics, $t s\tLf$, cannot be neglected (see also~\cite{Z1}). In order to derive the perturbation series in $s$ for $[e^{ t\LS}]_{\PS}$, we consider the Dyson expansion 
\begin{equation} 
[e^{ t\LS}]_{\PS}= \PS \,e^{ t [\LS]_{\PS}}\PS=\PS\left( e^{ t s\tLf}+\int_{0}^t\mathrm{d}u\, e^{(t-u) s\tLf}\,\delta \tLS\,e^{u [\LS]_{\PS}}\right)\PS, \label{eq:msregime1}
\end{equation}  
where  $\delta \tLS:=[\LS]_{\PS} - s\tLf= s^2 \tLs+\O(s^3 \lVert\tLt\rVert))$ of the order $\O(s^2)$ is treated as the perturbation to $s\tLf$ inside $\PS$. Using  $\PS=\Pz+s\Pf+\mathcal{O}(s^2)$ in Eq.~\eqref{eq:Ppert} and $e^{u [\LS]_{\PS}}\PS=\PS \,e^{u\LS}\,\PS$ we obtain
\begin{eqnarray} 
[e^{ t\LS}]_{\PS}&=&
\Pz \,e^{ t s\tLf}\, \Pz \,+\,s \left(\P^{(1)}\, e^{ t s\tLf}\, \Pz\,+\,\Pz\, e^{ t s\tLf}\,\P^{(1)}\right)\,+\,\O(s^2)+\label{eq:msregime2}\\
&&\,+\,\Pz\int_{0}^t\mathrm{d}u\, e^{(t-u) s\tLf}\,\delta \tLS\,\PS\,e^{u\LS}\Pz \,+\, \,t\,\O(s^3\,\lVert\tLs\rVert) ,\nonumber
\end{eqnarray} 
where the higher-order corrections are explained below. First, as both $s\tLf$ and $\LS$  are CPTP generators and $\lVert \T\rVert=1$ for $\T$ positive and trace-preserving~\cite{W05}, we have $\lVert e^{ ts\tLf} \rVert=\lVert e^{ t \LS } \rVert=1$ and $\lVert \Pz \rVert=1$. The first line in Eq.~\eqref{eq:msregime2} corresponds to $\PS\,e^{ t s\tLf}\,\PS$ and the higher-order corrections are of the order $\lVert\PS-\Pz-s \P^{(1)}\rVert\,\lVert\PS\rVert+s^2\lVert\P^{(1)}\rVert^2=\O(s^2)$ due to the norm $\lVert\cdot \rVert$ being submultiplicative, see reference [36] in the main text.  Furthermore, the corrections in the second line, which corresponds to the integral term in~\eqref{eq:msregime1}, are of the order 
\begin{eqnarray*}
\lVert \PS\rVert  \lVert \PS-\Pz\rVert  \lVert \int_{0}^t\mathrm{d}u\, e^{(t-u) s\tLf}\,\delta \tLS\,e^{u \LS }\rVert
&\leq& \lVert \PS\rVert  \lVert \PS-\Pz\rVert  \times t \,\lVert \delta \tLS\rVert \\
&=&\O(s) \times t\,(s^2\lVert \tLs\rVert+ \mathcal{O}(s^3\lVert \tLt\rVert))= t\,\O(s^3\lVert\tLs\rVert),
\end{eqnarray*}
where $\tLt$ is the third-order correction in $[\LS]_{\PS}$ (see Eq.~\eqref{eq:L3tilde}). 
Furthermore, since $\delta \tLS=s^2 \tLs+\O(s^3 \lVert\tLt\rVert))$ we also have 
\begin{equation*} 
\Pz\int_{0}^t\mathrm{d}u\, e^{(t-u) s\tLf}\,\delta \tLS\,\PS\,e^{u\LS}\Pz=s^2 \,\Pz\int_{0}^t\mathrm{d}u\, e^{(t-u) s\tLf}\,\tLs\,\PS\,e^{u\LS}\Pz+t\,\O(s^3\,\lVert \tLt\rVert)
\end{equation*} 
and further
\begin{equation*} 
s^2 \,\Pz\int_{0}^t\mathrm{d}u\, e^{(t-u) s\tLf}\,\tLs\,\left(\PS\,e^{u\LS}\right)\Pz
=s^2\int_{0}^t\mathrm{d}u\, e^{(t-u) s\tLf}\,\Pz\, \tLs\,{\Pz}\,e^{u s\tLf} \Pz+  t\, \O(s^3\,\lVert \tLs\rVert) + t^2\,\O(s^4\,\lVert \tLs\rVert^2),
\end{equation*} 
where we have used the Dyson expansion for $\PS\,e^{u [\LS]_{\PS}}=[e^{u\LS}]_{\PS}$, see Eq.~\eqref{eq:msregime1}, with corrections being the integral and the unitary evolution outside the SSM given by $\Pz$ (the first line in~\eqref{eq:msregime1}). Finally, we note that $\lVert\tLs\rVert=\mathcal{O}(\lVert\Ls\rVert\,+\,\lVert \Lf\rVert^2 \lVert\S\rVert )=\mathcal{O}(\lVert \Lf\rVert+\lVert\Ls\rVert)$ (cf.~Eq.~\eqref{eq:Lpert}), and $\lVert\tLt\rVert=\mathcal{O}(\lVert \Lf\rVert\,\lVert\Ls\rVert\,\lVert\S\rVert+\lVert \Lf\rVert^3 \lVert\S\rVert^2 )=\mathcal{O}(\lVert \Lf\rVert+\lVert\Ls\rVert)$ (cf. Eq. \eqref{eq:L3tilde})), which completes the proof of Eq.~\eqref{eq:msregime}. $\blacksquare$ \\


{\bf Timescale $\tau''(s)$}.  The metastable regime begins when the contribution from the fast decaying modes corresponding to the eigenvalues $\lambda_{m+1}(s)$, $\lambda_{m+2}(s)$, ..., becomes negligible and the initial relaxation to the $m$ low-lying modes takes place. The timescale $\tau''(s)$ in the decay of this contribution of the order  $\mathcal{O}\left( \left\lVert[e^{ t \LS}]_{\I-\PS}\right \rVert\right)$ is derived below as 
\begin{equation} 
\tau''(s)=\tau\,(1+\O(s )). \label{eq:taupp} 
\end{equation}

\noindent
\emph{Derivation}. Consider the Dyson expansion for $e^{t\LS}$ 
\begin{equation} 
e^{ t \LS}= e^{t\Lz}+\int_{0}^t\mathrm{d}u\, e^{(t-u) \Lz}\,\delta\LS\,e^{u \LS}, \label{eq:msregime3}.
\end{equation}  
where $\delta \LS:=\LS - \Lz=s\,\Lf+s^2\,\Ls$ is considered as a perturbation of $\Lz$, cf. Eq.~\eqref{eq:Lseries}.
As both $\LS$ and $\Lz$ are CPTP generators, we have $\lVert e^{t \LS} \rVert=\lVert e^{t \Lz} \rVert=1$~\cite{W05}. Using the expression \eqref{eq:Ppert} for $\mathcal{P}(s)$, we obtain
\begin{eqnarray}
\label{eq.semigroup.complement}
[e^{ t \LS}]_{\I-\PS}&=&(\I-\Pz)\, e^{t\Lz}\, (\I-\Pz)+ s\left[-(\I-\Pz)\,e^{ t \Lz} \,\Pf\,-\,\Pf\, e^{ t \Lz} \,(\I-\Pz) \right]\,+\,\O(s^2)\\
\nonumber
&&\,+\,(\I-\Pz)\,\int_{0}^t\mathrm{d}u\, e^{(t-u) \Lz}\,\delta \LS\,e^{u \LS}(\I-\Pz)+t\,\O(s^2  \,\lVert \Lf\rVert).
\end{eqnarray}
In the first line we used the multiplicativity of the norm, and 
$\lVert \PS-\Pz\lVert=\O(s)$. 
In the second line we bound the integral in Eq.~(\ref{eq:msregime3}) by $ t\, \lVert \delta \LS\rVert \leq t\,(s\lVert  \Lf\rVert+s^2\lVert  \Ls\rVert)$ we arrive at the correction $t\,\O(s^2  \,\lVert  \Lf\rVert)$. 

\vspace{1mm}

We now use the following definition of the relaxation time, $\tau$ as the shortest timescale such that for any initial state $\ri$, the system state relaxes to the stationary state as $\lVert e^{t\Lz} \ri-\Pz\ri \rVert\leq 2e^{-t/\tau} $, which implies 
$\lVert e^{t\Lz} (\I-\P) \Vert\leq 4 e^{-t/\tau}$. From Eq. \eqref{eq.semigroup.complement} we get
\begin{eqnarray}
\lVert [e^{ t \LS}]_{\I-\PS}\rVert&\leq& \lVert [e^{ t \Lz}]_{\I-\Pz}\rVert\,+\, 2 \,s \,\lVert [e^{ t \Lz}]_{\I-\Pz}\rVert\, \lVert \S\rVert \,\lVert \Lf\rVert  \,+\,\O(s^2)\nonumber\\
\nonumber
&&\,+\,2\int_{0}^t\mathrm{d}u\, \lVert [e^{ (t-u) \Lz}]_{\I-\Pz}\rVert\,\lVert \delta \LS\rVert\,+t\,\O(s^2  \,\lVert \Lf\rVert)
\nonumber\\
&\leq& 4e^{-t/\tau} (1\,+\, 2 \,s)  \,+\,8 s\,\tau \lVert \Lf \rVert\,+\,\O(s^2)\,+t\,\O(s^2  \,\lVert \Lf\rVert)\nonumber\\
&\leq&4 e^{-t/\tau} (1\,+\, 2 \,s)  \,+\,\O (s)+t\,\O(s^2  \,\lVert \Lf\rVert)
\end{eqnarray}
Note that the correction $t\,\O(s^2  \,\lVert \Lf\rVert)$ for times $t\ll\tau'(s)$ is of the same order as the leading corrections to the invariance of the MM, cf. Eq.~\eqref{eq:msregime}, and hence does not determine the timescale $\tau''(s)$ of the initial relaxation. Therefore, for times $\tau\ll t\ll \tau'(s)$, the contribution from the fast decaying modes  is a sum of terms of the order $\O (s)$ and of the same order as the corrections to the invariance of the MM. Similar results would be obtained for $\tau$ defined so that $\int_{0}^\tau\mathrm{d}t\, \lVert e^{ t \Lz}\,\rho_{\rm sup} -\Pz\rho_{\rm sup}\rVert =\sup_\ri \frac{1}{2}\int_{0}^\infty\mathrm{d}t\, \lVert e^{ t \Lz}\ri-\Pz\ri \rVert$, where $\rho_{\rm sup}$ is $\ri$ that gives the supremum.

Furthermore, in the situation when $\tau\sim (-\Re\lambda_{m+1})^{-1}$, i.e., when there are not too many modes contributing to the system dynamics, and $\lambda_{m+1}$ is non-degenerate, we have (see Eq. \eqref{eq:Lpert}) 
\begin{eqnarray*}
\tau''(s)&\sim&  (-\Re\lambda_{m+1}(s))^{-1}\sim (-\Re\lambda_{m+1})^{-1}\,+\,s\,\frac{\Re\,\Tr \,( \l_{m+1}\Lf \r_{m+1})}{(\Re\lambda_{m+1})^2} +\O (\tau\,s^2)\\
&\sim&\tau\left(1\,+\,s\,\tau\, \Re\,\Tr \,( \l_{m+1}\Lf \r_{m+1})\right)+\O (\tau\,s^2)\quad .
\blacksquare
\end{eqnarray*}
 
\bigskip

\noindent
{\bf Structure of the metastable manifold.} We consider now the projection of an evolved initial state onto the metastable manifold (MM) defined by  $\PS$.  Using the results \eqref{eq:msregime} and \eqref{eq:taupp} for the timescales $\tau'(s)$ and respectively $\tau''(s)$ we find that in the metastable regime $\tau''(s)\ll t\ll\tau'(s)$ the system state is approximated by (see Eq. (3) and (4) in the main text)
\begin{equation}
\rho_{MS}(t)
=\Pz\ri\,+\,s\left(-\S \Lf e^{ t s\tLf} \Pz-e^{ t s \tLf} \Pz  \Lf \S\right)\,+\,\O(s^2) \label{eq:msU}
\end{equation}
where the imaginary parts of the $m$ low-lying eigenvalues are given in the first order by the unitary dynamics $s\tLf$ within the MM (see Eqs.~\eqref{eq:Ppert} and~\eqref{eq:msregime}).  As $\Pz$ is the projection on the SSM of $\Lz$, $\rho_{MS}$ is approximately of the form given by Eq. (6) in the main text with the correction $\mathcal{O}(s \,\lVert S\rVert \lVert\Lf\rVert)=\mathcal{O}(s)$. Furthermore, this correction is of the same order as the corrections to invariance of the MM, i.e., the dissipative dynamics, for times $t=\O (s(\rVert \Lf\lVert+\rVert \Ls\lVert )^{-1}=\O (s^{-1}\tau)$ within the metastable regime $\tau''(s)\ll t\ll\tau'(s)$, see the second line in Eq.~\eqref{eq:msregime} and Eqs.~\eqref{eq:taup} and~\eqref{eq:taupp}. 

{\bf Coefficients of the MM}. Let us consider the generic case when the degeneracy of the first $m$ eigenvalues of $\Lz$ is lifted in the second-order perturbation theory. In this case the projections on the individual eigenmatrices are analytic and so are the coefficients of the MM, $(c_2(s),...,c_m(s))=(\Tr(\l_2(s)\ri),...,\Tr(\l_m(s)\ri))$.  Moreover, $\mathds{1}_\H$, $\l_2(0)$,..., $\l_m(0)$ and $\r_1(0)$, $\r_2(0)$, ..., $\r_m(0)$ correspond to the eigenbasis of  $\tLf+ \sum_{l=1}^{m''}\,\left[\tLs\right]_{\P_{l}}$ (cf. Eq.~\eqref{eq:Lpert2int}) and thus constitute a basis of the SSM of $\Lz$. Thus, for $s$ small enough, the set of coefficients representing the MM is well approximated by the image of an affine transformation of the $(m-1)$ degrees of freedom describing the SSM of $\Lz$, i.e. the $m$ entries of the DFS/NSS states, $p_l\omega_l$, $l=1,...m'$,  with the condition $\sum_{l=1}^{m'}p_l=1$ (cf. Eq. (6) of the main text). That affine transformation is determined by to the linear transformation between the basis of the SSM of the entries in $p_l\omega_l$, $l=1,...m'$ to $\r_1(0)$, $\r_2(0)$, ..., $\r_m(0)$. 

\emph{Derivation}. In the section on the spectrum of $\LS$, we argued that when the degeneracy is lifted in the second order, for $s$ small enough, the first $m$ eigenvalues of $\LS$ are analytic. Moreover, the eigenvalues are of the form $s\lambda_l^{(1)}+s^2\lambda_{l,j}^{(2)}+\mathcal{O}(s^3)$, where $\lambda_l^{(1)}$ is an eigenvalue of $\tLf$ with corresponding projection $\Pl$, and  $\lambda_{l,j}^{(2)}$ is an eigenvalue of $\Pl\tLs\Pl$. Due to the reduction process, the higher-order corrections correspond to the perturbation theory for  $s^{-1}(s^{-1}[\LS]_{\PlS}-\lambda^{(1)}_l\PlS)=[\tLs]_{\Pl}+s \tLt_{l} +\O(s^2)$ with the unperturbed operator $\tLf$ and an analytic perturbation. The first (third)-order perturbation $\tLt_{l}$ is given by Eq.~\eqref{eq:L3tildeL}. We thus have (cf. Eq.~\eqref{eq:Ppert2})
\begin{equation}
\PljS\,=\, \Plj\,+\, s \left( -\Plj\tLt_{l}\tS_{l,j}\,- \,\Plj\tLs\tS_l\,-\, \Plj\Lf \S+\,({\rm inv.})\right) \,+\,\O(s^2), \label{eq:Ppert3}
\end{equation}
where $P_{0,(l,j)}$ is the projection on the eigenmatrix corresponding to the eigenvalue $\lambda_{l,j}^{(2)}$ of $\Pl \tLs \Pl$, $\tS_{l,j}$ is the reduced resolvent for $\Pl \tLs \Pl$ at $\lambda_{l,j}^{(2)}$ restricted to $\Pl$, and $\tS_l$ is the reduced resolvent for $\tLf$ at $\lambda^{(1)}_l$ restricted to $\P_0$. Note that the corrections depend via $\tS_l$ and $\tS_{l,j}$ on the way the degeneracy is lifted inside the SSM in the first and the second order of the perturbation theory. The right eigenvector corresponding to $\PljS$ is thus \emph{proportional} to 
\begin{equation*}
\l_{l,j}(s)\,\propto\, \l_{l,j}\PljS  = \l_{l,j} \,-\,s\,\left(\l_{l,j}\tLt_{l}\tS_{l,j}\,- \,\l_{l,j} \tLs \tS_l\,-\, \l_{l,j}\Lf \S\right)\,+\,   \mathcal{O}(s^2),
\end{equation*}
where  $ \l_{l,j}$ is the left eigenmatrix of $\Pl \tLs \Pl$ corresponding to $\lambda_{l,j}^{(2)}$. Note that since the projection $\PljS$ is of rank 1, the eigenmatrix $\l_{l,j}$ can be replaced by any matrix $\l$ such that  $\l\PljS\neq0$. Let us assume $\l_{l,j}$ is Hermitian (see the paragraph with Eq.~(4) in the main text), so that the coefficient $c_{l,j}=\Tr(\l_{l,j}\ri)$ is real.  

Consider $\l_{l,j}(s)$ normalised in the \emph{spectral norm} $\lVert \l_{l,j}(s) \rVert_{\infty}=\max_{|\psi\rangle\in\H,\langle\psi|\psi\rangle=1}|\langle\psi|\l_{l,j}(s)|\psi\rangle|$, which corresponds to the maximal absolute value of the $\l_{l,j}(s)$ eigenvalues. Note that  $\lVert \l_{l,j}(s) \rVert_{\infty}= \max_{\ri} |\Tr(\l_{l,j}(s)\ri)| =\max_{\ri} |c_{l,j}(s)|$. From the Hermitian perturbation theory for $\l_{l,j}(s)$, the eigenvalues of $\l_{l,j}(s)$ are analytic~\cite{K95}, but $\lVert \l_{l,j}(s) \rVert_{\infty}$ does not have to be differentiable at $s=0$, which happens when  the extreme eigenvalues of $\l_{l,j}$ obey $|c_{l,j}^{\max}|=|c_{l,j}^{\min}|$. Nevertheless, for a given sign of $s$, $\lVert \l_{l,j}(s) \rVert_{\infty}$ is analytic for $s$ small enough. Therefore, we arrive at
\begin{equation}
c_{l,j}(s)=\frac{\Tr(\l_{l,j}(s)\ri)}{\lVert \l_{l,j}(s) \rVert_{\infty}}\,=\,  c_{l,j} (1 - s \,c_{l,j}^{{\rm ex},(1)})  \,-\,s\,\Tr\left[\left(\l_{l,j}\tLt_{l}\tS_{l,j}\,- \,\l_{l,j}\tLs\tS_l\,-\, \l_{l,j}\Lf \S\right)\ri\right]\,+\,   \mathcal{O}(s^2),
\end{equation}
where we assumed $\lVert \l_{l,j} \rVert_{\infty}=1$ and $c_{l,j}^{{\rm ex},(1)}$ related to the first-order correction to $c_{l,j}^{\min}$ or $c_{l,j}^{\max}$ with its sign depending on the sign of $s$.  Therefore, for $s$ small enough the set of coefficients representing the MM is simply an affine transformation of the degrees of freedom of the SSM of $\Lz$ as given in Eq. (6) of the main text.

Consider an alternative case in which the coefficient $c_{l,j}(s)$ is ''normalised"  by the difference of the extreme eigenvalues of $\l_{l,j}(s)$ ,  $\Delta c_{l,j}(s):=  c_{l,j}^{\max}(s)-c_{l,j}^{\min}(s)$.  This  ''normalisation" is convenient as the range of all coefficients determining the MM is of the same length $1$, which is also  the case for probabilities in a simplex or a Bloch ball, see Fig.~2. in the main text.  From the Hermitian perturbation theory for $\l_{l,j}(s)$ we have that $\Delta c_{l,j}(s)$ is analytic in $s$ and thus
\begin{equation}
c_{l,j}(s)=\frac{\Tr(\l_{l,j}(s)\ri)}{\Delta c_{l,j}(s) }\,=\,  c_{l,j} ( 1 - s \,(\Delta c_{l,j}^{(1)})^{-1})  \,-\,s\,\Tr\left[\left(\l_{l,j}\tLt_{l}\tS_{l,j}\,- \,\l_{l,j}\tLs\tS_l\,-\, \l_{l,j}\Lf \S\right)\ri\right]\,+\,   \mathcal{O}(s^2)
\end{equation}
where we assumed $\Delta c_{l,j}(0)=1$ and $\Delta c_{l,j}^{(1)}$ is the difference between first-order corrections in  $c_{l,j}^{\max}(s)$ and  $c_{l,j}^{\min}(s)$. $\blacksquare$

\bigskip

\noindent
{\bf Effective dynamics in the metastable manifold.} Previously we showed that the dynamics in the metastable regime is approximated by unitary transformation of the MM with generator $\tLf$. Here we show that for times 
$\tau'(s) \leq t\ll  s^{-1} \,\tau'(s)=s^{-3}\O(\tau)$ (i.e. following the metastable regime) the dynamics in the MM is dissipative and is characterised by the CPTP generator $\tLS:=s\tLf+  s^2 [\tLs]_\Pz$ on the SSM of $\Lz$. In particular, we prove that 
\begin{equation}
[e^{ t \LS}]_{\PS}=  e^{ t\tLS} \Pz +s\left(-\S \tLf e^{t \tLS} \Pz-e^{t \tLS} \Pz  \tLf \S\right)\,+\,\O(s^2)\,+\,t\,\O(s^3 \tau^{-1}).\label{eq:mmdynamics}
\end{equation}
We note that dynamics generated in the SSM by $[\tLs]_{\Pz}$ was previously discussed in~\cite{Z3} for the special case of a Hamiltonian perturbation (see Eq.~\eqref{eq:Lseries}) and $\tLf=0$.

\emph{Proof}. Note that from the fact that  $\LS$ is trace-preserving, it follows that it features  the left eigenmatrix $\mathds{1}_\H$ corresponding to $0$-eigenvalue, which, by construction, also holds true for $[\LS]_{\PS}$. Therefore, $\L_{\rm eff}$ is trace-preserving.

From Eq.~\eqref{eq:Lpert} we write $\L_{\rm eff} = \widetilde{\mathcal{L}}(s)+ \Delta\LS $, with $\Delta \LS$ regarded as a perturbation whose size is in general $ \lVert\Delta \LS\rVert = \O(s^2 \, \lVert  \tLf \rVert)$, while in the case when  $\tLf=0$ we have $ \lVert\Delta \LS\rVert = \O(s^3 \, (\lVert\Lf\rVert+\lVert\Ls\rVert))$, see the third-order correction for $[\LS]_\PS$ in Eq.~\eqref{eq:L3tilde}.
%
%
The Dyson expansion for $\L_{\rm eff}$ with $\Delta\LS$ as the perturbation is
$$
[e^{ t \LS}]_{\PS}=\PS\, e^{t  \widetilde{\mathcal{L}}(s)} \PS + \PS \int_{0}^t\mathrm{d}u\, e^{(t-u)\widetilde{\mathcal{L}}(s)}\,\Delta \LS\,\PS\,e^{u \LS} \PS,
$$
where we used $e^{u \L_{\rm eff}} \PS=\PS\,e^{u \LS} \PS$. We further have 
\begin{eqnarray*}
[e^{ t \LS}]_{\PS}&=&\Pz\, e^{t \widetilde{\mathcal{L}}(s)} \Pz +s\left(-\S \tLf e^{t \widetilde{\mathcal{L}}(s)} \Pz-e^{t \widetilde{\mathcal{L}}(s)} \Pz  \tLf \S\right)\,+\,\O(s^2)\,+\\
&&\,+\, \Pz \int_{0}^t\mathrm{d}u\, e^{(t-u) \tLS}\,\Pz \,\Delta \LS\,\Pz\,e^{u  \LS} \Pz\,+\, t\, \O(s^3 \,\lVert\tLf\rVert  \,+\,s^4 \, (\lVert\Lf\rVert+\lVert\Ls\rVert)), 
\end{eqnarray*}
%
Note that $\lVert e^{t \tLS} \rVert\ = \| e^{t \tLS} \mathcal{P} + (\mathcal{I}-\mathcal{P})\|\leq 3 $ since $\tLS$ is a CPTP generator on $\mathcal{P}$. Due to submultiplicativity of the norm, the corrections to $\Pz\, e^{t\widetilde{\mathcal{L}}(s)} \Pz$ in the first line are $\mathcal{O}(s^2 (\lVert  \Lf \rVert\,\lVert\S \rVert)^2)=\mathcal{O}(s^2)$. In the second line corresponding to the integral term, the corrections are bounded by $ t\, \O(s\, \lVert  \Lf \rVert\,\lVert\S \rVert \,\lVert\Delta \LS\rVert )=t\, \O(s^3 \,\lVert\tLf\rVert  \,+\,s^4 \, (\lVert\Lf\rVert+\lVert\Ls\rVert))$.  
From Eq. \eqref{eq:Lpert} we get $\lVert[\Delta \LS]_\Pz\rVert= \O(s^3 \, (\lVert\Lf\rVert+\lVert\Ls\rVert))=\O(s^3 \tau^{-1})$ which implies that the leading correction in the second line is $\O(s^3 \tau^{-1})$. $\blacksquare$
%

{\bf Stationary state}. We note that a stationary state of the dynamics perturbed away from the degeneracy have been studied in~\cite{B08}. In the case when the degeneracy is lifted in the second order of the perturbation theory, we have that (cf. Eq.~\eqref{eq:Ppert3})
\begin{equation} 
\rss(s)\,=\,  \rss \,+\,s\,\left( \tS_{1,1}\,\tLt_{1}\,\rss  \,- \,\tS_1\, \tLs \,\rss\,-\, \S \,\Lf \,\rss \right) \,+\,\O(s^2),
\end{equation} 
where $\rss$ is the unique stationary state of the generator in Eq.~\eqref{eq:Lpert2int}. Let $\P_{1}$ denote the projection on the ($\lambda_1^{(1)}=0$)-eigenspace of $\tLf$. $\tS_1$ is the reduced resolvent of $\tLf$ at $0$, restricted to $\Pz$, $\tLt_{1}$ is the third-order perturbation in the reduction process for eigenspace $\P_{1}$, see Eq. \eqref{eq:L3tildeL}, and $\tS_{1,1}$ is the resolvent of $[\tLs]_{\P_{1}}$, restricted to $\P_{1}$. Note that from the orthonormality of the eigenbasis  of the CPTP generator $\tLf+\sum_{l=1}^{m''}\,\left[\tLs\right]_{\P_{l}}$ (the first and second order of perturbation theory for $[\LS]_\PS$), we further have $\Tr\,((\I-\P_{1,1})\r)=0$ for any matrix $\r$, where $\P_{1,1}(\cdot)=\rss \Tr(\cdot)$, and thus $\Tr\,\rss (s)=1+\O(s^2)$.  


\bigskip
\noindent
{\bf Proof of the CPTP property of the effective generator.} We now prove that $[\tLs]_\Pz$ and $\tLf$ generate CPTP dynamics on the SSM given by $\Pz$. We use Theorem 3.17 from~\cite{D80} on convergence of one-parameter semigroups, whose statement we recall here for the special case of finite dimensional spaces. Let $\Z(s)$, $\Z$ be generators of one-parameter semigroups $\T^t(s):=e^{t \Z(s)}$,  $\T^t:=e^{t \Z}$ on a Banach space $\mathcal{B}$, and assume 
that for each $X$ in a spanning set of  $\mathcal{B}$ there exist $X(s) \in \mathcal{B}$ such that $\lim_{s\rightarrow 0} X(s)=X$ and $\lim_{s\rightarrow 0}\Z(s)(X(s))=\Z(X)$. Then  for all $T$ the limit $\lim_{s\rightarrow 0} \sup_{t\leq T} \lVert \T^t(s)(X)-\T^t(X)\rVert=0$, where $\lVert\cdot\rVert$ is the norm in $\mathcal{B}$.


\vspace{1mm}

\emph{Proof for $[\tLs]_\Pz$.} To prove the CPTP property consider $|\psi\rangle=\frac{1}{\sqrt{D}}\sum_{i=1}^D|e_i\rangle\otimes|e_i\rangle \in \H\otimes\H$, where $\{|e_i\rangle\}_{i=1}^D$ is an orthonormal basis of the system space $\H$. We choose $X=\left(\Pz\otimes\I\right)\,(|\psi\rangle\langle \psi|)\in \mathcal{B}(\H\otimes\H)$ and $\Z=[\tLs]_{\Pz}\otimes\I$ so that  $M_t:=\T^t (X)$ is the Choi matrix for $e^{\tLs} \Pz$. By choosing appropriate CPTP generators $\Z(s)$  and matrices $X_s$ we will show that $M_t$ is a limit of Choi matrices of quantum channels. Thus for all $t$, $M_{t}$ is positive and $\Tr_1 (M_t) =D^{-1} I_{\H}$, where $\Tr_1$ denotes the partial trace over the first subsystem in $\H\otimes\H$, and consequently $\tLs$ generates CPTP dynamics on the SSM given by $\Pz$. To prove this, we choose $\Z(s)=s^{-2}(\LS-s[\Lf]_{\Pz})\otimes\I $, which is a CPTP generator on $\H\otimes\H$ as $[\Lf]_{\Pz}$ is a generator of unitary quantum dynamics. By defining $X(s)=X+s X^{(1)}+s^2 X^{(2)}$, where $X^{(1)}=-\left( \S \Lf \otimes\I \right)\,X$ and $X^{(2)}=\left(\S\Lf \S\Lf \otimes\I \right)\,X-\left(\S\Ls\otimes\I \right)\, X$, we arrive at the conditions of the theorem 3.17 in~\cite{D80} with the norm $\lVert\cdot\rVert$ being the trace norm (see [36] in the main text). We note that the generator property of $[\tLs]_{\Pz}$ was previously discussed in~\cite{Z3} for the special case of the Hamiltonian perturbation (see Eq.~\eqref{eq:Lseries}) and $\tLf=0$. $\blacksquare$

\emph{Proof for $\tLf$.} Similarly, to prove that $\tLf=[\Lf]_\Pz$ generates CPTP dynamics on the SSM given by $\Pz$, we need to choose  $X=\left(\Pz\otimes\I\right)\,(|\psi\rangle\langle \psi|)$ and $\Z=\tLf\otimes\I$. By considering $\Z(s)=s^{-1}\LS\otimes\I$ and  $X(s)=X-s\left( \S \Lf \otimes\I \right)\,X$ we arrive at the conditions of the theorem 3.17 in~\cite{D80}. We note that $\tLf$ was proven to be a unitary generator in~\cite{Z1,Z2}. $\blacksquare$
 
\bigskip
\noindent
{\bf Expressions for higher-order corrections $\tLt$ and $\tLt_l$.} We have that $[\LS]_\PS=s\tLf+s^2\tLs+s^3\tLt+\O(s^4(\lVert\Lf\rVert+\lVert\Ls\rVert))$, where $\tLf$ and $\tLs$ are given in Eq.~\eqref{eq:Lpert} and the third-order correction is~\cite{K95}
\begin{eqnarray}
\,\tLt&=& \,-\,\Pz\Lf \Pz \Ls \S \,-\, \Pz \Ls \Pz \Lf \S \,-\, \Pz \Lf \S \Ls\Pz \,-\, \Pz \Ls \S\Lf \Pz\,-\, \S \Lf \Pz \Ls \Pz \,-\, \S \Ls \Pz \Lf \Pz \,+\nonumber\\
&&\, +\,  \Pz \Lf \Pz \Lf \S \Lf \S \,+\, \Pz \Lf \S \Lf \Pz \Lf \S \, +\, \Pz \Lf \S \Lf \S \Lf \Pz\,+\,\nonumber\\
&&\,+\, \S \Lf \Pz \Lf \Pz \Lf \S\,+\, \S \Lf \Pz \Lf \S \Lf \Pz \, +\,  \S \Lf \S \Lf \Pz \Lf \Pz\,- \nonumber\\
&&\,-\, \Pz \Lf \Pz \Lf \Pz \Lf \S^2\, -\, \Pz \Lf \Pz \Lf \S^2 \Lf \Pz\, -\,  \Pz \Lf \S^2 \Lf \Pz \Lf \Pz \,-\, \S^2 \Lf \Pz \Lf \Pz \Lf \Pz,  \label{eq:L3tilde}\\
\, [\tLt]_\Pz&=&\,-\,\Pz \Lf \S \Ls\Pz  \,-\, \Pz \Ls \S\Lf \Pz\, +\, \Pz \Lf \S \Lf \S \Lf \Pz\,- \nonumber\\
&&\, -\, \Pz \Lf \Pz \Lf \S^2 \Lf \Pz\, -\,  \Pz \Lf \S^2 \Lf \Pz \Lf \Pz. 
\end{eqnarray}
Due to reduction process for $[\LS]_\PS$ we further obtain that $[\LS]_{\Pl(s)}=s\lambda_l^{(1)}\Pl(s)+s^2[\tLs]_\Pl+s^3\tLt_l+\O(s^4)$, where $\Pl(s)$ is a projection on the $\lambda_l^{(1)}$-group with $\lambda_l^{(1)}$ being an eigenvalue of $\tLf$ and 
\begin{eqnarray}
\tLt_l&=& [\tLt]_{\P_{l}}\,+\,\lambda_l^{(1)}\,\P_{l} \Lf \S^2 \Lf \P_{l} \,-\,\P_{l} \tLs \tS_l \tLs \P_{l}\, -\nonumber\\
&&\,-\,\P_{l}  \tLs \P_{l}  \tLs \tS_l\,-\,\tS_l \tLs \P_{l}  \tLs\P_{l} \,-\,\P_{l}  \tLs \P_{l}  \Lf \S \,-\,\S \Lf \P_{l}  \tLs\P_{l}, \quad l=1,...,m''. \label{eq:L3tildeL}
\end{eqnarray}


\subsection*{Comment on the conjecture for Class B}

For class B a proof of our conjecture of the MM structure appears difficult.

The convex analysis tools used in the classical proof \cite{g1,g2,g3,g4} cannot be used for the quantum case as they rely on the finite number $D$ of pure states of a finitely-dimensional classical system.  Note, however, that by using any tools of convex analysis for the MM represented by the set of coefficients $(c_2,\ldots,c_m)$, and exploiting (approximate) positivity of the metastable states, one could at most prove the structure of fixed points of positive (cf. completely positive) maps \cite{I13}, which is richer than Eq.~(6).  For example, for $m=3$ there can exist non-commuting eMS ($2\times2$ real Hermitian matrices) in contrast to $m\geq4$ for a smallest DFS/NSS of a qubit. In order to exploit complete positivity one would need to work with the dynamics extended to $e^{t\L}\otimes \I_{\mathcal{B(H)}}$, which has $m\times D^2$ low-lying eigenvalues and thus the simplicity of the geometric representation of the MM is lost.

For the stronger conjecture determining also the structure of the metastable states, the difficulty lies in the fact that the existing proof of the SSM structure relies on the property that eigenmatrices corresponding to 
strictly zero eigenvalue of a CPTP generator (or the eigenvalue 1 of a CPTP quantum channel) form a von-Neumann algebra and thus are of the form given in Eq.~(6) of the main text~\cite{B08,W}.  We cannot rely on the algebra structure of metastable states as for states with approximately the block structure, this structure will not be preserved with the same approximation for products of them. This corresponds to  the corrections to complete positivity of the dynamics (of the same order as the corrections to the stationarity, cf.\ Eq.~(3)) being progressively accumulated with each multiplication of the eigenmatrices (cf. the proof of the SSM structure in~\cite{W}). It is likely one could use a proof such as in~\cite{R12} deriving Eq.\ (6) by exploiting properties of a projection on steady states without its multiple applications. 


These points will be elaborated further in \cite{M}.


\end{document}